\documentclass[doublespacing]{elsart}  

\usepackage[usenames]{color}
\usepackage{amsmath,amssymb,amsfonts}
\usepackage{latexsym}
\usepackage{textcomp}
\usepackage{pifont}
\usepackage{dcolumn}
\usepackage{hhline}
\usepackage{rotating}
\usepackage{multirow}
\usepackage{tipa}
\usepackage{ae,aecompl}
\usepackage{mathrsfs}
\usepackage{colortbl}
\usepackage{longtable}
\usepackage{lineno}
\usepackage{framed}
\usepackage{subfigure}

\usepackage{icarus}

\newlength\jataille
\newcommand{\figgauche}[3]%
{\jataille=\textwidth\advance\jataille by -#1
\advance\jataille by -.5cm
\begin{minipage}[c]{#1}
   \includegraphics[width=#1]{#2}
\end{minipage}\hfill
\begin{minipage}[c]{\jataille}
   \footnotesize #3 \normalsize
\end{minipage}}

\usepackage{natbib}

\newcommand{\BibTeX}{ \textrm{B\kern-.05em\textsc{i\kern-.025em b}\kern-.08em
    T\kern-.1667em\lower.7ex\hbox{E}\kern-.125emX} }

\begin{document}

\begin{frontmatter}
\begin{center}
\title{The opposition effect in Saturn's rings seen by Cassini/ISS: \\
I. Morphology of phase curves} 

%
%
%
%

\author[deau]{Estelle D\'eau}, 
\author[deau]{S\'ebastien Charnoz},
\author[dones]{Luke Dones},
\author[deau]{Andr\'e Brahic}, and
\author[porco]{Carolyn C. Porco}

\address[deau]{CEA Saclay DSM/Irfu/SAp AIM UMR 7158, Orme des Merisiers, Bat. 709 91191 Gif-sur-Yvette FRANCE}
\address[dones]{Southwest Research Institute, 1050 Walnut Street, Suite 300 Boulder CO 80302, USA}
\address[porco]{CICLOPS, 3100 Marine Street Suite A353, Boulder CO 80303, USA}

\end{center}


\vspace{1cm}
Number of pages: \pageref{lastpage} \\
Number of tables: \ref{lasttable}\\
Number of figures: \ref{lastfig}\\
\end{frontmatter}

\newpage
\noindent Running head: The opposition effect in Saturn's rings seen by Cassini/ISS 1. 

\noindent Direct editorial correspondence to~: \\
Estelle \textsc{D\'eau} \\
CEA Saclay \\
Service d'Astrophysique \\
AIM, UMR 7158 \\
Orme des Merisiers Bat. 709 \\
91191 Gif-sur-Yvette FRANCE \\
phone: +33 1 69 08 80 56 \\
fax: +33 1 69 08 65 77 \\
e-mail : estelle.deau@cea.fr \\

\newpage
\begin{linenumbers}
\begin{abstract}

The Cassini cameras have captured the opposition effect in Saturn's
rings with a high radial resolution at phase angles down to
0.01\textsuperscript{o} in the entire main ring system. We derive
phase functions from 0.01\textsuperscript{o} to 25\textsuperscript{o} of phase angle in the Wide-Angle Camera
clear filters with a central wavelength
$\lambda_{\textrm{cl}}$=0.611~$\mu$m and phase functions from 0.001\textsuperscript{o} to 25\textsuperscript{o} of phase angle in 
the Narrow-Angle and Wide-Angle Cameras color filters (from the blue, $\lambda_{\textrm{bl}}$=0.451~$\mu$m to the 
near infrared $\lambda_{\textrm{it}}$=0.752~$\mu$m). We characterize the morphology 
of the phase functions of different features in the
main rings. We find that the shape of the phase function is accurately
represented by a logarithmic model (Bobrov 1970, in Surfaces 
and Interiors of Planets and Satellites, Academic, edited by A. Dollfus). For practical
purposes, we also parametrize the phase curves by a simple
linear-by-part model (Lumme and Irvine 1976, Astronomical Journal, 81, p865), 
which provides three morphological
parameters~: the amplitude and the Half-Width at Half-Maximum (HWHM)
of the surge, and the slope S of the linear-part of the phase function
at larger phase angles. Our analysis demonstrates that all of these
parameters show trends with the optical depth of the rings. These trends imply that 
the optical depth is a key-element determining the physical properties which 
act on the opposition effect. Wavelength variations of the morphological parameters of the surge 
show important trends with the optical depth in the green filter ($\lambda_{\textrm{gr}}$=0.568~$\mu$m), 
which implies that grain size effects are maximum in this wavelength. 

\end{abstract}




\newpage
\section{Introduction} \label{intro} 


When one views the rings of Saturn with the Sun directly behind the
observer, a phenomenon called the \textit{opposition effect} can be
seen. The opposition effect, also known as the opposition surge, is a
sudden, nonlinear rise in brightness with decreasing phase (Sun--ring--observer) 
angle that occurs as the phase angle approaches zero. The opposition effect,
which was observed for most Solar System bodies 
\citep[e.g.,][]{1997Icar..128....2H,1999Icar..141..132S,2000Icar..147...94B,2003Icar..166..276B,2004Icar..172..149S,2005Icar..173...66V,2007Sci...315..815V,2006JQSRT.100..325R} 
and \citep{2007AJ....133...26R} was
discovered in the course of M\"uller's long-term photometry of the
Saturn system, beginning in 1878 \citep[e.g.][]{1885AN....110..225M,1893POPot..30..198M}. 
\citet{1884AN....109..305S,S1887} inferred that the opposition effect was
due to the rings, since Jupiter did not show a comparable opposition
brightening \citep{1975SSRv...18....3P}. 

Most recent studies of the opposition effect in the Saturn's rings were based on ground based and spatial data which 
resolved the main rings \citep{1976AJ.....81..865L,1979AJ.....84.1408E,2002Icar..158..224P,2007PASP..119..623F}. 
Earth-based observations, though valuable, have a low to moderate spatial resolution \citep[for the Hubble
Space Telescope, 1 pixel at Saturn =285~km and the full-width at
half-maximum (FWHM) of the point-spread function of Hubble's Camera =7~pixels, 
see][]{2002Icar..158..199C}. This indicates that
the signal is averaged over large regions in the rings. Unfortunately,
the rings are highly heterogeneous features that may present rapid
spatial variations of their optical properties \citep{1987ApJS...63..749E}. 
So the interest of \textit{spacecraft} observations is the
ability to probe the signal in very narrow ring features ($\thicksim$40~km 
in the present paper), which may have much more uniform optical
properties. This allows an easier study of the opposition effect which
is an already complex phenomenon. \\
Pioneer~11, Voyager~1 and Voyager~2 did not observe the rings at phase angles 
near zero. However, Cassini, which became an artificial satellite of Saturn 
in 2004, is the first spacecraft to observe the opposition effect in the rings,
with several instruments \citep{2006LPI....37.1461N,2007Icar..191..691A}, 
including the Narrow Angle (NAC) and Wide Angle (WAC) cameras of the
Imaging Science Subsystem (ISS). For our data set (Table\,\ref{table_oe_info_images}),
typically for the WAC 1~pixel=40~km and the FWHM is 1.8~pixels, 
for the NAC 1~pixel=5~km and the FWHM is 1.3~pixels 
\citep{2004SSRv..115..363P}. With Cassini images we are thus able to characterize how
the surge varies throughout individual features in the ring system.

The opposition effect is now known to be the combined effect of the coherent 
backscatter (at small phase angles) which is a constructive interference of photons in a medium of using 
grains of wavelength size and the shadow hiding (at larger phase angles) which consists of shadows cast by particles  
on other particles that become invisible to the observer \citep{1997Icar..128....2H}. 
Some analytical models have tried to combined these two effects 
\citep{1999Icar..141..132S,2002Icar..157..523H}, but recent laboratory 
measurements and data analysis show that models' results could be compromised by 
some physical assumptions on the scatterer elements, the wavelength-dependence and the the angular size of 
the source \citep{2007JGRE..11203001S,2007JQSRT.106..487S,2008LPI....39.1498D}. 
Two other effects, the external near-field effect and the single scattering 
internal-field coherence, were never been observed solely 
because they are intrinsically inseparable from the coherent backscatter effect and are for the 
moment only numerically simulated \citep{2007Icar..188..233P,2007JQSRT.106..360M}.   

Both shadow-hiding and coherent backscattering are likely to play
roles in determining the shape of the phase curves of Solar System bodies at low
phase angles \citep{1997Icar..128....2H,1998Icar..133...89H}. Shadow-hiding 
probably dominates at phase angles greater
than a few degrees, while coherent backscattering takes effect at the
very smallest phase angles. The shadow-hiding effect gives clues about
the three-dimensional structure of a layer of ring particles which, according 
to \citet{1974IAUS...65..441K}, could have a typical size of $\bar{r}=15$m 
\citep[see also][]{2003Icar..164..428S,2007PASP..119..623F}. 
By contrast, the coherent backscattering component sheds light on the nature of the 
ring particle surfaces at a scale not much larger than optical wavelengths. 
Indeed, previous photometric studies which investigate the opposition effect 
in the Saturn's rings determined ranges of typical size for coherent-backscattering 
effect of about $d=10~\mu$m  \citep{1992MNRAS.254P..15M,2002Icar..158..224P,2007PASP..119..623F}.
Thus, the shape of the opposition effect provides constraints on features
vastly smaller than the camera's resolution. \newline
Like \citet{2002Icar..158..224P}, we refer to particles as macroscopic and individual objects 
in the rings and grains as microscopic objects at the surface of rings' particles. 
Using this terminology, ring particles might be covered by regolithic grains. 

This is the first of a series of papers dealing with the opposition
effect in Saturn's rings as seen by ISS/Cassini. In the present paper, we focus on the
characterization of the morphological modeling of the shape of opposition effect in the
rings. Since the recent theories have difficulties to
link the behavior of the opposition effect with the physical
properties of the surface material \citep{2000Icar..147..545N,2007JGRE..11203001S}, 
we think it is necessary to have an experimental approach, consisting in (1) deriving the shape's parameters and (2) trying to 
find correlations among themselves, and with the optical depth. 
In the second paper (D\'eau et al. in preparation), we will use recent analytical photometric 
models to derive some of the physical properties of the ring material. Indeed, to constrain completely the physical properties of the ring material, photometric and polarimetric phase curves are needed, however, Cassi/ISS did not captured the opposition spot with its polarized filters. 
\newline
In section\,\ref{observe_et_reduit} of the present paper, we describe the ISS/Cassini data set and our procedure for
extracting photometric data from images and fitting empirical
models to the data. In section\,\ref{results}, we characterize the morphology of the opposition surge at 
different locations in the main rings and focus our attention on
cross-correlations among the morphological parameters and correlations of these 
parameters with the optical depth and the wavelength. Finally, in section\,\ref{discussion} 
we discuss these results and examine to which extents photometric models may explain the 
morphological trends derived in section\,\ref{observe_et_reduit}.


\section{Observations and reductions} \label{observe_et_reduit} 

\subsection{The Cassini Imaging Data Set} \label{data_set} 

The Cassini ISS instrument is composed of two cameras, a wide angle
camera (WAC) and a narrow angle camera (NAC) equipped with
1024$\times$1024 CCD matrices. Both use a set of about twenty filters
ranging from near-infrared to ultraviolet \citep{2004SSRv..115..363P}. \newline
Here, we focus only on filters in the optical domain, with first the blue, green, red and near infrared filters, 
that will be called hereafter COLOR filters. According to \citep{2004SSRv..115..363P}, COLOR filters from NAC and WAC cameras 
do not have exactly the same central wavelength. For the blue filters, the NAC can have two different combinations of filters (CL1,BL1) or (BL1,CL2) which lead to two central wavelengths~: $\lambda_{\textrm{bl}}^{\textrm{\textsc{nac}}}=0.440\mu$m and $\lambda_{\textrm{bl}}^{\textrm{\textsc{nac}}}=0.451\mu$m. In contrast, for the WAC, the blue (CL1,BL1) filter is characterized by a wavelength of about $\lambda_{\textrm{bl}}^{\textrm{\textsc{wac}}}=0.460\mu$m. Because the spectral width of these filters is about $\pm$0.050$\mu$m, we consider all images from blue filters of the NAC and the WAC as a consistent whole. The green filter (CL1,GRN) has almost the same spectral characteristics for the two cameras since the central wavelength of $\lambda_{\textrm{grn}}^{\textrm{\textsc{nac}}}=0.568\mu$m for the NAC and $\lambda_{\textrm{grn}}^{\textrm{\textsc{wac}}}=0.567\mu$m for the WAC. The red filter for the NAC corresponds to the filter (RED,CL2) with a central wavelength of $\lambda_{\textrm{red}}^{\textrm{\textsc{nac}}}=0.650\mu$m. For the WAC, the combination (CL1,RED) is at a central wavelength of  $\lambda_{\textrm{red}}^{\textrm{\textsc{wac}}}=0.649\mu$m, which does not change to the that of the NAC at one nanometer. Finally, the filter (CL1,IR1) in the near infrared shows a difference of 10 nanometers between the central wavelength of the NAC ($\lambda_{\textrm{ir}}^{\textrm{\textsc{nac}}}=0.752\mu$m) and that of the WAC ($\lambda_{\textrm{ir}}^{\textrm{\textsc{wac}}}=0.742\mu$m). In summary, with this moderately high spectral resolution, the combination of images that are not coming exactly from the same filters are responsible for a shift of the central wavelength from 0.001 to 0.010 micron. Because each filter has a non negligible spectral range ($\pm$0.050$\mu$m), we consider without distinction the COLOR filters from the NAC and WAC cameras and in the rest of the paper, we take as a reference the following wavelengths ($\lambda_{\textrm{bl}}=0.451\mu$m; $\lambda_{\textrm{grn}}=0.568\mu$m; $\lambda_{\textrm{red}}=0.650\mu$m and $\lambda_{\textrm{ir}}=0.752\mu$m) to designate each COLOR phase curve. \newline
The other part of images discussed in the present paper has been taken in CLEAR mode of the WAC,
designating the absence of filters resulting in a spectral bandwidth
spanning from 0.20 to 1.10~$\mu$m (the central wavelength is $\lambda_{\textrm{cl}}=0.611\mu$m and the spectral range is $\pm$0.450$\mu$m). \newline
For the present paper, the selected observation campaign of the opposition effect on the Saturn's rings 
is divided in four sequences from May 2005 to July 2006 (Table\,\ref{table_oe_info_images}). 

\textbf{[Table\,\ref{table_oe_info_images}]}

\begin{figure}[!ht]
\begin{center}
\includegraphics[width=8cm]{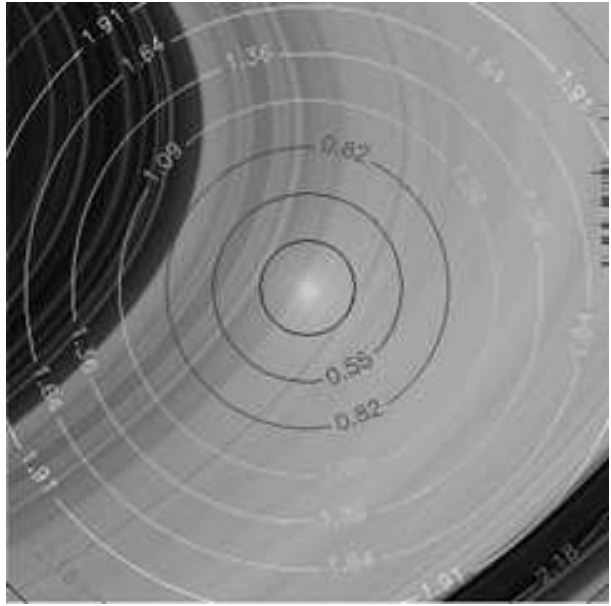}
\caption{\label{represent_oe_images} The opposition effect in the
B~ring. A typical image of the \textit{26 June} sequence captured by
the Wide Angle Camera (W1498453136.IMG). Concentric circles correspond to identical phase curves 
computed in the image (isophase). Numbers in the concentric circles represent the solar phase angles (in degrees).}
\end{center}
\end{figure}

\begin{figure}[!ht]
\begin{center}
\includegraphics[width=8cm]{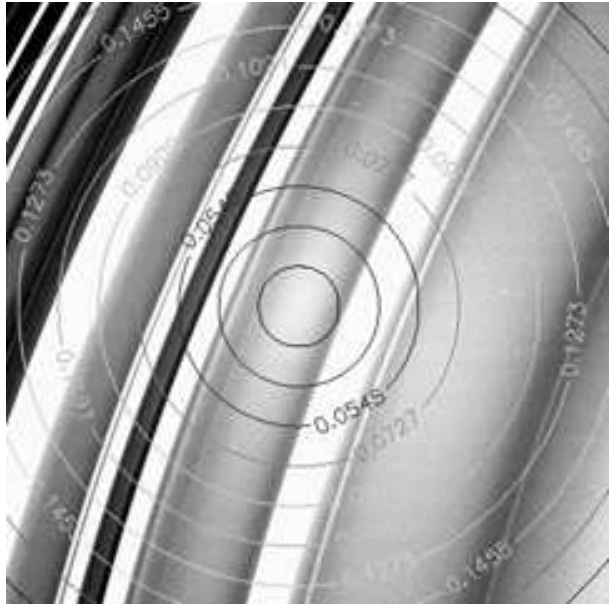}
\caption{\label{oe_hisurge1} The opposition effect in the C~ring. A
typical image of the \textit{20 may} sequence captured by the Narrow
Angle Camera (N1595278165.IMG). The contrast is enhanced to make more
visible the opposition spot in the C~ring. Concentric circles correspond to identical phase curves 
computed in the image (isophase). Numbers in the concentric circles represent the solar phase angles (in degrees).}
\end{center} 
\end{figure}

These observations were conduced conjointly by the VIMS, CIRS
and ISS instruments. Three of ISS sequences were obtained using the WAC
only and the last one using the BOTSIM mode (using both cameras simultaneously). 
Depending on the sequence, different filters were used and different spatial and angular
resolutions were achieved. In the present paper, the first of a series
aimed at a detailed study of the opposition effect, we will focus on
sequences obtained using the WAC camera in CLEAR filter mode and sequences using NAC and WAC 
cameras in COLOR filter mode, Table\,\ref{table_oe_info_images}. Images of the B and C~ring are shown
in Fig.\,\ref{represent_oe_images} and \ref{oe_hisurge1} to illustrate the quality of the data set. 

\textbf{[Fig.\,\ref{represent_oe_images}, Fig.\,\ref{oe_hisurge1}]}

\begin{figure}[!ht]
\includegraphics[width=16cm]{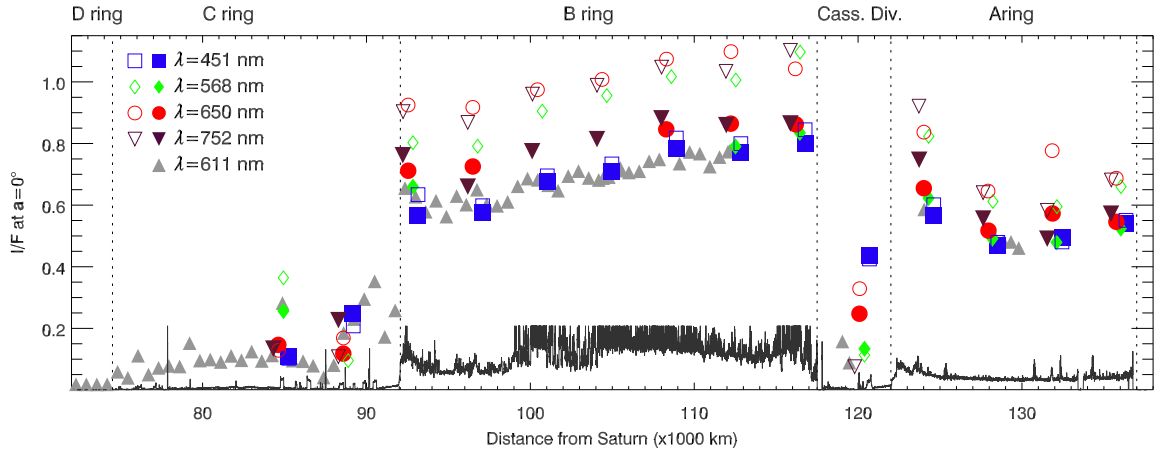}
\caption{\label{oe_images_color_location} Radial location of the opposition spot in 
the images taken in CLEAR and COLOR filters (filled symbols represent WAC images and empty symbols represent NAC images). 
We give in the $y$-axis the normalized brightness $I/F$ of the minimum phase angle in the image, the 
$x$-axis is the corresponding distance from Saturn of this point. The vertical dotted lines correspond 
to ring boundaries. The optical depth of PPS/Voyager 
is plotted as a radial reference.}
\end{figure}

In the CLEAR filter mode, the wide angle camera captured rings at zero phase angle 
two times in 2005, on June 7 and June 26 (they will be designated as \textit{June~7}
and \textit{June~26} sequences in the rest of the paper, for practical
purpose). \newline
In the \textit{June~7} sequence, the WAC images have an
average resolution of $\thicksim$~44~km per pixel; the rings are observed
in reflexion. In each individual image, the phase angle varies by
about 3 degrees. In the majority of the set, the opposition point at
zero phase angle is visible with a very good sampling~: about
0.005 degree per pixel. This set covers the full ring system by
tracking the opposition spot. Most of the set has an excellent
photometric quality, however in dark regions the strong Saturnshine produces ghost images of the
secondary mirror perturbing the signal significantly. The amplitude of
this artifact is quite constant, about 0.05 in I/F \citep[I/F is the phase-corrected reflectivity, see][]{2004SSRv..115..363P}, which does not
significantly affect the photometry of the bright regions as A and
B~rings. However it is very troublesome for dark regions like the
D~ring, the C~ring and the Cassini Division. This is why images from the
\textit{June~7} sequence have not been considered for these two
regions. The same instrument artifact has been observed and discussed
in \citep{2007Icar..188...89H}.  \newline 
The \textit{June~26} sequence
has similar characteristics, see Table\,\ref{table_oe_info_images}, 
with a better radial resolution of about 30~km per pixel and spans the B and C~rings. The
presence of the ghost signal in the C~ring is not clear. A detailed
photometric analysis in these images
(cf. section\,\ref{phase_curves_oe}) shows that the derived phase
function is consistent with an unperturbed signal, very
differently from images in which the ghost were clearly identified. It
is why the images of the C~ring taken on \textit{June~26} have been considered
for the C~ring phase function. \newline 
The \textit{July 23} sequence has
the best radial resolution ($\thicksim$13~km.pixel$^{-1}$) and spans
all A, B, C~rings and the Cassini Division (the radial location
of the opposition spot in each image is represented in the
Fig.\,\ref{oe_images_color_location}). In this set, the ghost artifact
is absent owing to a much larger angular separation with the bright
Saturn globe. Some images of this sequence provide also the larger phase angles.

\textbf{[Fig.\,\ref{oe_images_color_location}]}

In COLOR filters mode, the ISS instrument uses the BOTSIM mode in the 
\textit{May~20} sequence, using both cameras simultaneously for a ride in A and B rings, 
Cassini Division and the outer C ring (Fig.\,\ref{oe_images_color_location}). 
However, because the boresight of the NAC and the NAC are not exactly the same, 
NAC images did not represent a zoom of the opposition spot in WAC images. 
Thus, WAC and NAC images captured at the same time have opposition spots that are not exactly at 
the same location in the rings. The radial resolution in NAC images is 4.6~km.pixel$^{-1}$ and 
spans from 44 to 66~km.pixel$^{-1}$ for WAC images in COLOR filters. 
The other sequences (\textit{Dec~31}, \textit{Feb~20} and \textit{Apr~25}), 
did not have the opposition spot but provided the larger phase angles.

\subsection{Data reduction} \label{data_reduction}

\subsubsection{Calibration}
Raw images are calibrated first with a standard pipeline described in \citep{2004SSRv..115..363P} and called
CISSCAL, using the 3.4 version. The DN values are converted into the I/F units. 
This ratio is dimensionless, with $I$ is the specific
intensity measured by Cassini, and $\pi F$ is the incident flux from
the Sun. So it is a measure of the local reflectivity under the current
observing geometry. \newline
Cassini images are not directly exploitable to study the photometric
behavior of rings. In order to do this, the relevant data we need are 
the so-called \textit{phase function} which is linked to the I/F ratio
as a function of the phase angle, and corrected from effects of
observation geometry. A full procedure has been designed to reconstruct the
phase function from different images with different observation
geometry and resolutions and is detailed below.

\subsubsection{Extraction procedure}  \label{extraction}
The first step is to reproject the images in a (Radius, Longitude)
frame, in which features at a same radius from Saturn are horizontally
aligned. This procedure critically depends on the quality on the
navigation. When possible, the edges of the A, B, C rings as well as
ring features reported in \citep{1987ApJS...63..749E} were used as
fiducial references. Distances reported in \citep{1987ApJS...63..749E} 
were corrected according to \citep{1990AJ....100.1339N}. 
The resulting navigation error on WAC images varies from
1 to 1.5 pixels from the center to the edge of the image. \newline The I/F
in a ring at constant distance from Saturn is obtained by extracting
the data on a line of constant radius (\textit{i.e.} an horizontal
line) in the reprojected map of calibrated brightness. Other
geometrical parameters are extracted in the same way~: phase angle
$\alpha$, cosine of incidence angle $\mu_{0}$, cosine of the emission
angle $\mu$, optical depth of the rings obtained from the PPS Voyager
instrument ($\lambda=0.260~\mu$m), \citep{1987ApJS...63..749E};
radius scale has been corrected with the procedure of \citet{1990AJ....100.1339N}. \newline 
This procedure works very well for structures with width larger than the navigation accuracy. 
However radial structures are visible at all scales down to one pixel,
(Fig.\,\ref{oe_hisurge1}). For structures radially smaller than
5~pixels, navigation errors make false the extraction along a line of constant radius 
close to the edges of the image, inducing accidental
extraction in nearby different features. To overcome this problem and
to ensure that we always extract the same ring feature, we developed a
\textit{ring tracking} technique using a basic pattern recognition
algorithm to follow a single feature. Extensive visual check of the
result shows the method is reliable down to 1~pixel of radial width.

\subsubsection{Construction of the phase function}  \label{build_curve}
The ultimate information we need is the phase function of individual ring particles 
to characterize their surface properties. Unfortunately, the signal from an individual particle is
heavily altered because of the finite thickness of the rings \citep{PW2007}. Also,
the signal's intensity depends on observation angle with respect to the ring's  
normal. Inverting such complex collective photometric effects would require the 
use a detailed light scattering code with many assumptions concerning the photometric
properties of particles.  Such code has been developed by 
\citet{1992Natur.359..619S,1995Icar..117..287S,1994MNRAS.269..493R,1999DPS....31.4403P,PW2007}  and \citet{2007PASP..119..623F}. However, for our present purposes, they cannot used to derive the phase functions in hundreds of different regions as 
we wish to do here. Consequently, as a first approximation, we use in 
the present paper the classical approach of \citet{1960ratr.book.....C} 
linking the I/F to the
phase function with the following assumptions~: homogeneous layer of 
particles and single scattering. This latter assumption may be justified 
for phase angles smaller than $\thicksim$30\,\textsuperscript{o} \citep{2002Icar..158..199C}. 
In reflexion, the phase function $\varpi_{0}\cdot P(\alpha)$ is derived from the solution to the
radiative transfer equation (the designation \textit{phase function} is not strictly 
accurate since what we really determine is the product of the single 
scattering albedo, $\varpi_{0}$, times the particle phase function, $P(\alpha)$)~:

\begin{equation}
\varpi_{0}\cdot
P(\alpha)=\frac{I}{F}\times\frac{4(\mu+\mu_{0})}{\mu_{0}}\times\left(
1 - e^{-\tau\left(\frac{1}{\mu}+\frac{1}{\mu_{0}}\right)} \right)^{-1}
\label{eq_chandrasekhar}
\end{equation}

with $\tau, \mu,\mu_{0}, \alpha$ standing for~: the normal optical  depth, 
cosine of emission angle, cosine of incidence angle and phase
angle respectively.  In order to allow future comparisons with
detailed numerical models, numerous analytical parametrization of the
observations are provided by morphological model in
section\,\ref{describe_model_oe}. As one can see, the value of the
optical depth is necessary to derive the phase function. 
Preceding works has shown that the exponential factor can be
neglected in first approximation for Earth-based observations 
\citep{2002Icar..158..199C,2002Icar..158..224P}.  However, we
noticed that we obtained much more coherent results when taking into
account the exponential factor, when comparing results from different
geometry of observations. Consequently, we keep the initial formula of
\citet{1960ratr.book.....C}, as previous photometric studies based on
spacecraft observations of Saturn's rings \citep{1989Icar...80..104D,1991PhDT.........8C,F1992,SHOW1992,dones1993}.

$P(\alpha)$ obtained with Eq.\,\ref{eq_chandrasekhar} is the 
particle's disk integrated phase function which determines the angular 
distribution of single scattered radiation from the body as a whole. The phase function is
normalized over the solid angle~$\Omega$ to the single scattering albedo~: 
$\varpi_{0}=\frac{1}{4\pi} \int P(\alpha)~\textrm{d}\Omega$. 
To derive the albedo, the full phase curve, from 0 to 180 degrees of
phase angle, must be known. In this study, we have restricted our data
(0<$\alpha$<25 degrees), so that we may avoid separating the albedo
from the phase function. The derivation of the albedo will be presented in 
the second paper which relates ISS data from 0 to 180 degrees.

\subsubsection{Finite size of the Sun} \label{solar_size_fini_sat}
\begin{figure}[!ht]
\hspace{0.5cm}
\includegraphics[width=7.5cm]{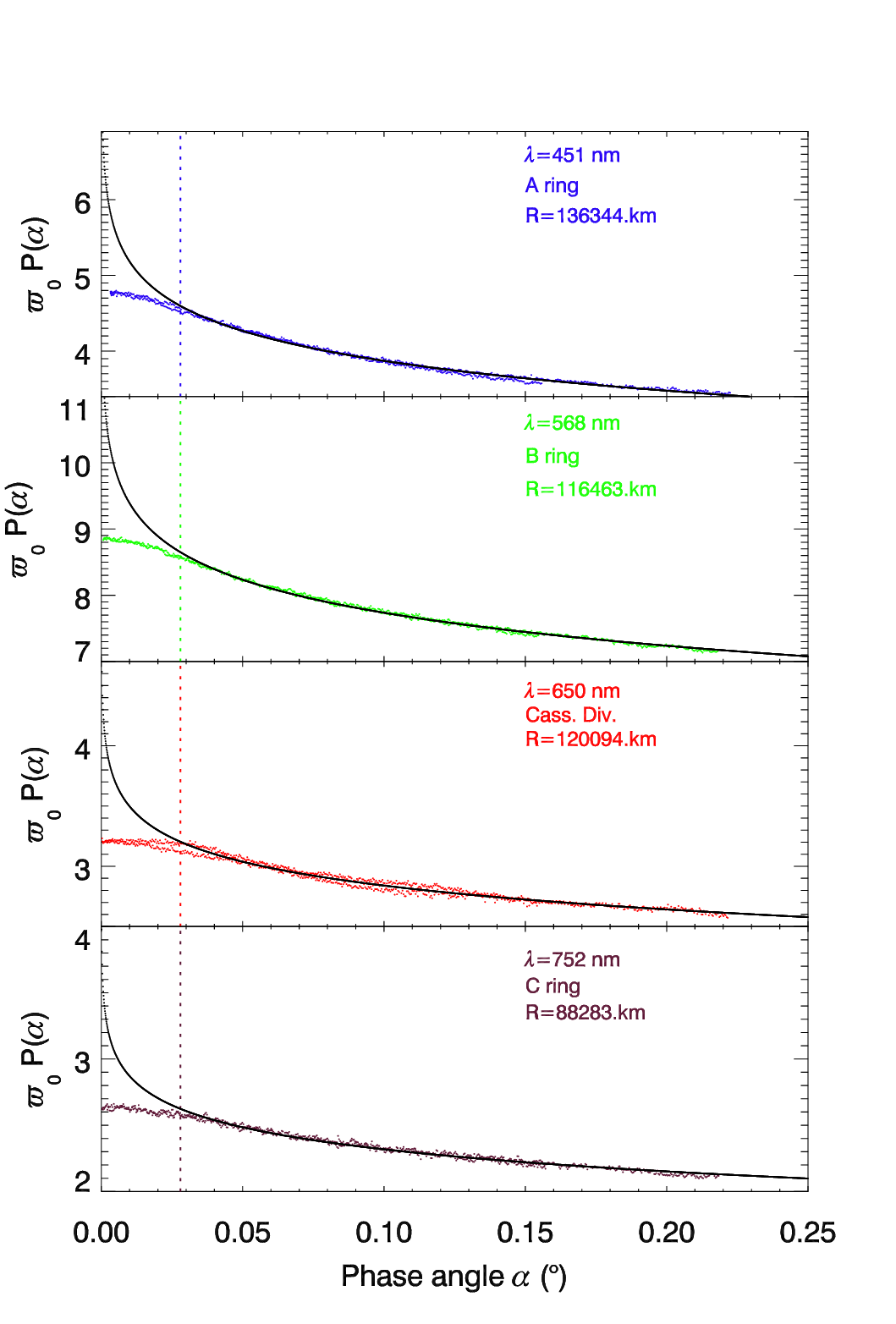}
\includegraphics[width=7.5cm]{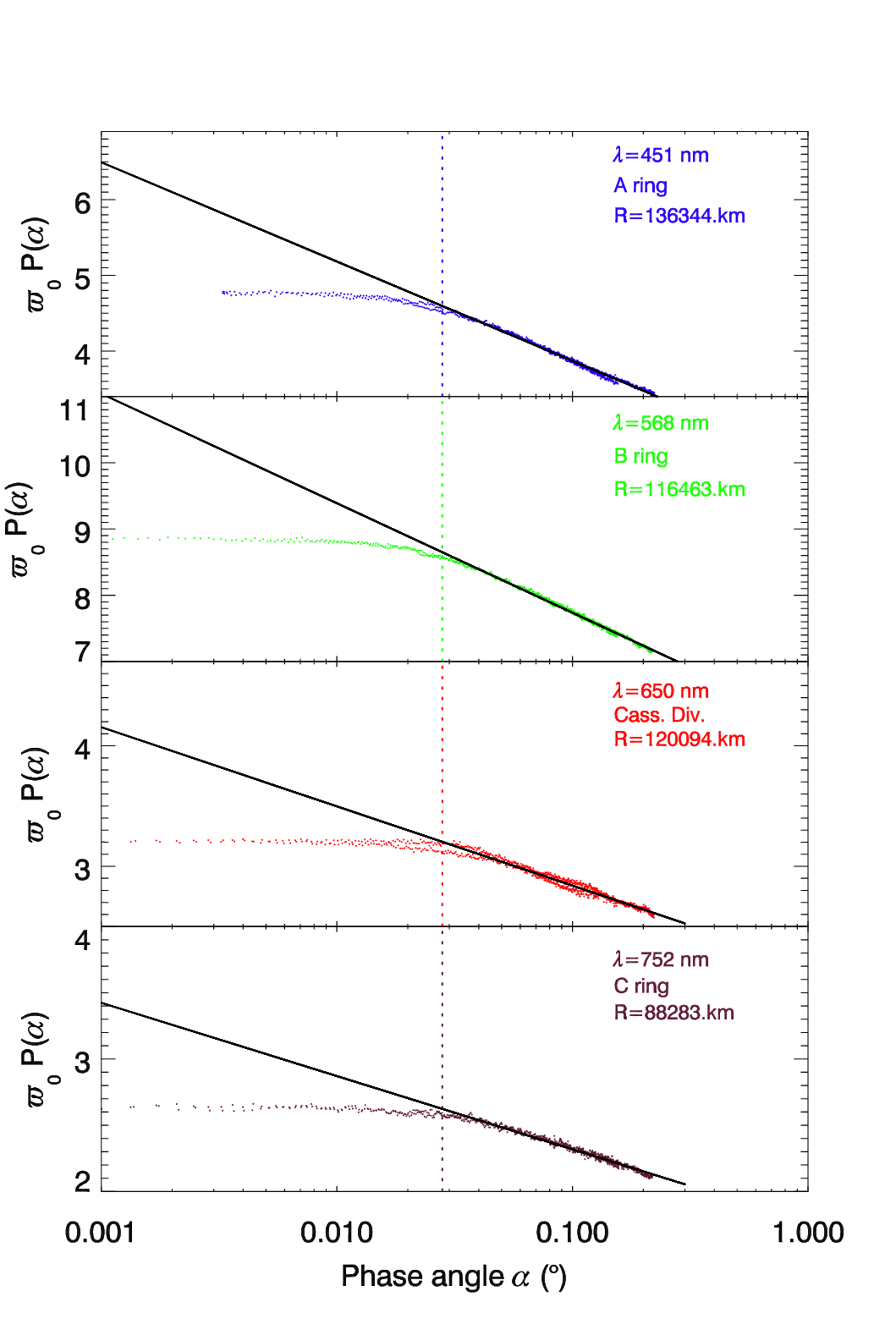}
\caption{\label{represent_nac_all_rings} Extracted COLOR phase curve from NAC images of the \textit{May 20} 
sequence in A, B, C rings and Cassini Division in linear scale (a) and logarithmic scale (b). 
The vertical dotted line in (a) and (b) corresponds to the angular size of the Sun. 
The solid curves in (a) and (b) correspond to the logarithmic model of \citet{1970sips.conf..376B}.}
\end{figure}

Generally speaking, all phase curves present a bright opposition surge
below 1\textsuperscript{o} and a slope decreasing linearly for
$\alpha>$1\textsuperscript{o}. Whereas WAC images 
go down to $\thicksim$0.01\textsuperscript{o} of phase angle (Fig.\,\ref{represent_oe_images}), NAC
images taken in COLOR filters capture the opposition spot at better 
angular resolution (Fig.\,\ref{oe_hisurge1}), the resulting phase functions 
from NAC images go down to $\thicksim$0.001\textsuperscript{o} (Fig.\,\ref{represent_nac_all_rings}). 
This is the first time the opposition spot is imaged at such fine scale. \newline
Note that we define the phase angle as the angle between the vector pointing to the Sun's center, 
and to the spacecraft, from the observed point. So from a strictly \textit{mathematical point of view} 
there is no lower limit to the phase angle value despite the source's finite size. For an extended illumination source, 
the phase curve should be the integral of a point source phase function over 
the Sun angular radius. Thus on a more \textit{physical point of view} 
it is not possible to observe the phase function at angle below the Sun's angular radius \citep{1974IAUS...65..441K,1991AVest..25...71S}. \newline
Fig.\,\ref{represent_nac_all_rings} demonstrates that the 
opposition surge flattens (in all rings and at all wavelengths) at phase angles below 0.029\textsuperscript{o}, in good agreement with the Sun's angular radius at the date of observations \citep[0.0291~degree, given by 
$\alpha_{\odot\textrm{min}}=\arcsin\frac{r_{\odot}}{R_{\odot-\textrm{Saturn}}}$ 
where $r_{\odot}=6.96\times10^{5}$~km is the radius of the Sun 
and $R_{\odot-\textrm{Saturn}}$ is the heliocentric distance of Saturn][]{2000ssd..book.....M}. 

\textbf{[Fig.\,\ref{represent_nac_all_rings}]}

\subsection{Data fit with Morphological models}  \label{describe_model_oe}
The purpose of the present paper is to provide an accurate description
of the morphological behavior of the observed phase curves. This is
the very first step prior to any attempt of further modeling, which will be the subject of a forthcoming paper. 
As a consequence, special care has been given to parametrize the phase functions 
conveniently. In addition morphological parametrization is necessary
to compare efficiently hundreds of phase curves at different locations
in the rings and derive statistical trend as will be done in
section\,\ref{results}. \newline
Several morphological models have been used in
the past to describe the shape of the phase functions~:
the logarithmic model of \citet{1970sips.conf..376B}, the linear-by-part model of
\citet{1976AJ.....81..865L} and the linear-exponential model of
\citet{2001JQSRT..70..529K}. The specific properties of these
three models make them adapted for different and complementary
purposes. The logarithmic model is interesting for direct comparisons
with numerical models, the linear-by-part model is convenient to
describe the shape in an intuitive way, and finally the
linear-exponential model is adapted for comparison with other studies previously published.

\subsubsection{The logarithmic model} As \citet{1970sips.conf..376B}, 
\citet{1976AJ.....81..865L} and \citet{1979AJ.....84.1408E}, we remark that a
logarithmic model describes very well the solar phase curves of the Saturn's rings. It depends on
two parameters ($a_{0}$ and $a_{1}$). This first morphological model has the following form~:

\begin{equation}
\varpi_{0}P(\alpha)=a_{0} + a_{1}\times\log(\alpha)
\end{equation}

In general, this model is the best morphological fit to the data. It
is reasonably accurate down to 0.029\textsuperscript{o} of phase angle. 
This is due to the finite angular size of the Sun which flattens data below 0.029\textsuperscript{o}, 
whereas the logarithmic function continues increasing (see section\,\ref{solar_size_fini_sat} and 
Fig.\,\ref{represent_nac_all_rings}a). For large phase angles, the fit is
satisfactory up to $\alpha\simeq$15 degrees, this was also noticed by \citet{2007Icar..191..691A} 
who fitted temperature phase curves of the C ring with CIRS/Cassini.  \newline
Whereas the meaning of
$a_{0}$ and $a_{1}$ are not easily interpretable in term of shape, to allow an easy
comparison with future numerical simulations the values of
these two parameters are reported in Table~1 of Electronic supplement material.

\subsubsection{The linear-exponential model} 
This model describes the shape of the phase function as a combination of an
exponential peak and a linear part. Its main interest is that it has
been used in previous work for the study of the backscattering of
Solar System's icy satellites and rings \citep{2001JQSRT..70..529K,2002Icar..158..224P,2007PASP..119..623F}. 
We give the details of this model~: the 4~parameters are the intensity of the peak~$I_{p}$, 
the intensity of the background~$I_{b}$, the absolute slope of the linear part~$I_{s}$ 
and the width of the exponential~$w$, such that the phase function is represented by~:

\begin{equation}
\varpi_{0}P(\alpha)=I_{b} + I_{s}\cdot\alpha + I_{p}\cdot
e^{-\frac{\alpha}{2w}}
\end{equation}

With these 4~parameters, we characterize the shape of the phase function by introducing three
morphological parameters~: A, HWHM and S designating the amplitude of
the surge, the half-width at half-maximum of the surge and the absolute slope
at large phase angles respectively, so that~:

\begin{equation}
\textrm{A}=\frac{I_{p}+I_{b}}{I_{b}}~~~~~~~~\textrm{HWHM}=2\cdot
\ln2w~~~~\textrm{and}~~~~\textrm{S}=-I_{s}
\end{equation}

As \citet{2007PASP..119..623F}, we noticed that this model did 
not fit well the phase curves, in particular, the derived model parameters appeared to depend substantially on the phase angle coverage (see section\,\ref{incomplete_data_fits}), preventing a robust comparison of data. This is due to the fact that we did not use the converging procedure of \citet{2001JQSRT..70..529K} but rather the common downhill minimization technique, as done by the previous users of the linear-exponential model \citep{2002Icar..158..224P,2007PASP..119..623F}. Moreover, it seems that the angular scale at which the phase function is observed may strongly influence the model parameters. So to avoid this problem, we used a much simpler morphological model that appeared much more robust for the comparison of heterogeneous data set : the linear-by-part model of \citet{1976AJ.....81..865L}, also we found a much better match of the data with the linear-by-part model of \citet{1976AJ.....81..865L} (see below). \newline
However, the linear-exponential was considered in the present study only to understand variations of A, HWHM and S between the \citep{2002Icar..158..224P,2007PASP..119..623F} phase curves and our phase curves (see section\,\ref{incomplete_data_fits} and section\,\ref{compare_hst_iss}).

\subsubsection{The linear-by-part model} 
For an intuitive description of the main features of the phase curves, the
linear-by-part model is the most convenient one. It is constituted of
two linear functions fitting both the surge at small phase
angle ($\alpha<\alpha_{1}$) and the linear regime at higher phase
angle ($\alpha>\alpha_{2}$)~:

\begin{eqnarray}
\varpi_{0}P(\alpha<\alpha_{1})=-A_{0}\times\alpha + B_{0} \newline
\varpi_{0}P(\alpha>\alpha_{2})=-A_{1}\times\alpha + B_{1}
\end{eqnarray}

\citet{1976AJ.....81..865L} and \citet{1979AJ.....84.1408E} use
$\alpha_{1}$=0.27\textsuperscript{o} and
$\alpha_{2}$=1.5\textsuperscript{o}. However, we encountered difficulties with the value of $\alpha_{1}$.
By testing several values of $\alpha_{1}$, it appears that for our data set, values of $\alpha_{1}$
less than 0.3\textsuperscript{o} yield a general overestimation of $A_{0}$, especially in
the C~ring and values of $\alpha_{1}$ greater than 0.3\textsuperscript{o} yield an
underestimation of $A_{0}$ only in the B~ring. Consequently we found
the our data were better reproduced using
$\alpha_{1}=0.3$\textsuperscript{o} which is now adopted in the rest
of the paper. \newline With this four outputs~: the two absolute slopes~$A_{0}$ and $A_{1}$ and the
two y-intercepts~$B_{0}$ and $B_{1}$, the shape of the curve is characterized by 
the morphological parameters A, HWHM and S are then defined by~:

\begin{equation}
\textrm{A}=\frac{B_{0}}{B_{1}}~~~~~~~~\textrm{HWHM}=\frac{(B_{0}-B_{1})}{2(A_{0}-A_{1})}
~~~~\textrm{and}~~~~\textrm{S}=A_{1}
\end{equation}

The purpose of this model is not, of course, an accurate description
of the data but rather a convenient description of the main trends of
the phase curve.

\subsubsection{Stability of the morphological parameters} \label{incomplete_data_fits}
\begin{figure}[!ht]
\begin{center}
\includegraphics[width=16cm]{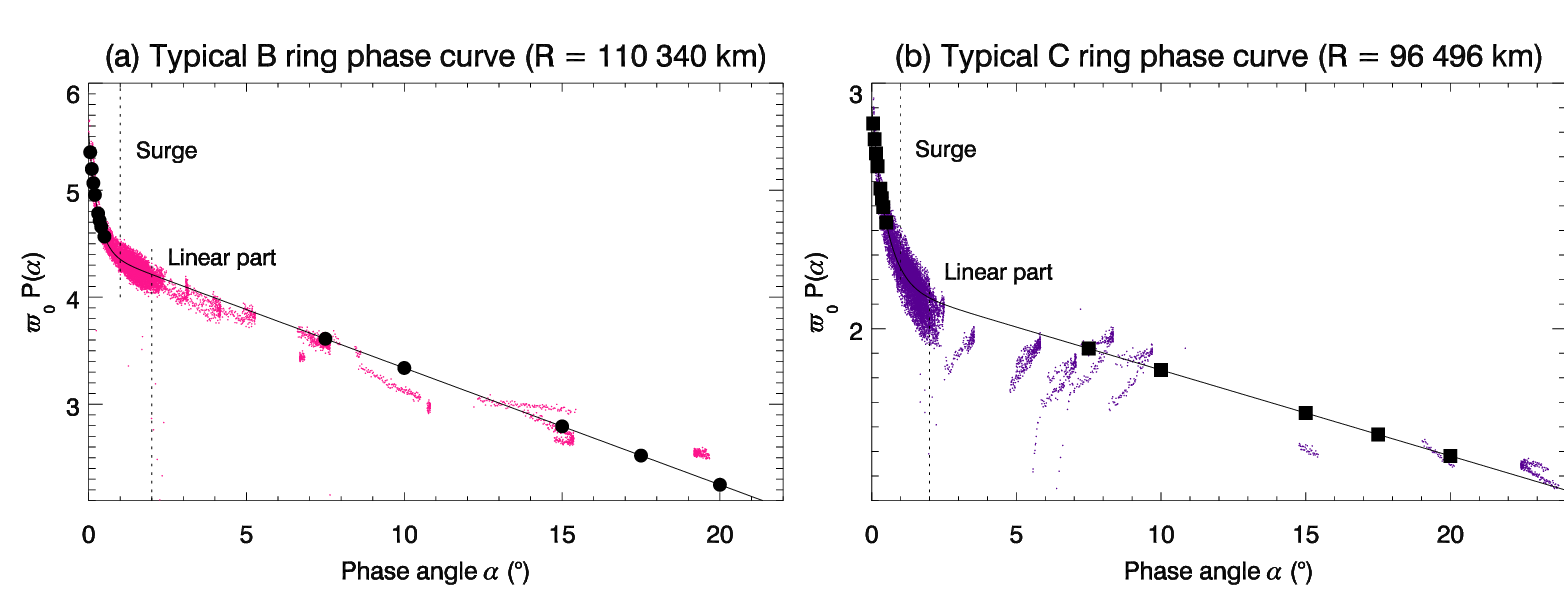}
\caption{\label{fig2} Typical phase curves of the B~ring (a) and C~ring (b) in CLEAR filters chosen for testing the stability of 
linear-exponential model of \citet{2001JQSRT..70..529K} when the phase angle coverage is incomplete. 
Circles and squares correspond to the cutoff of the incomplete phase function and the solid curves correspond to the initial fit of 
the linear-exponential model of \citet{2001JQSRT..70..529K}. The vertical dotted lines correspond to boundaries of the surge domain
and the linear part domain that we have tested (see Fig.\,\ref{fig3}).}
\end{center}
\end{figure}

In order to compare properly our results with those of \citet{2002Icar..158..224P} 
and \citet{2007PASP..119..623F}, for which the Saturn's rings phase curves did not 
have the same phase angle coverage, we have tested the influence of the portion 
0.05\textsuperscript{o}-0.4\textsuperscript{o} and 6\textsuperscript{o}-25\textsuperscript{o} 
on the converging solution. Indeed, phase curves of \citet{2002Icar..158..224P} do not 
have data under 0.3\textsuperscript{o}, and both studies of \citet{2002Icar..158..224P} and 
\citet{2007PASP..119..623F} do not have data over 6.5\textsuperscript{o} \citep[the maximum phase angle reached from Earth is about 
$\alpha_{\odot\textrm{max}}=\arcsin\frac{R_{\odot-\textrm{Earth}}}{R_{\odot-\textrm{Saturn}}}\lesssim6.5$\textsuperscript{o} 
where $R_{\odot-\textrm{Earth}}$ is the heliocentric distance of the Earth and 
$R_{\odot-\textrm{Saturn}}$ is the heliocentric distance of Saturn computed with orbits of][]{2000ssd..book.....M}. \newline
With two typical Saturn's rings phase curves of ISS (Fig.\,\ref{fig2}), 
we have removed data by section of 0.1\textsuperscript{o} and fit the pseudo-incomplete 
phase curve with the linear-exponential model, which provides the new solution, designated 
by $A_{\textrm{remove}}$, HWHM$_{\textrm{remove}}$ and $S_{\textrm{remove}}$. The initial 
solution found for fuller phase function (0.01\textsuperscript{o}-25\textsuperscript{o}) is called $A_{\textrm{optimal}}$, 
HWHM$_{\textrm{optimal}}$ and $S_{\textrm{optimal}}$. Both solutions are obtained with a downhill minimization technique, to reproduce the fitting method used by \citet{2002Icar..158..224P} and \citet{2007PASP..119..623F}. It is wise to recall that  
\citet{2001JQSRT..70..529K} proposed a converging procedure more accurate, however because \citep[e.g.][]{2002Icar..158..224P,2007PASP..119..623F} did not used it, and because we want to reproduce not the best converging solutions but the deviations lead by the fits driven by \citet{2002Icar..158..224P} and \citet{2007PASP..119..623F}, that is why we use the downhill method and not the probability distribution method of \citet{2001JQSRT..70..529K}.

\begin{figure}[!ht]
\begin{center}
\includegraphics[width=7.5cm]{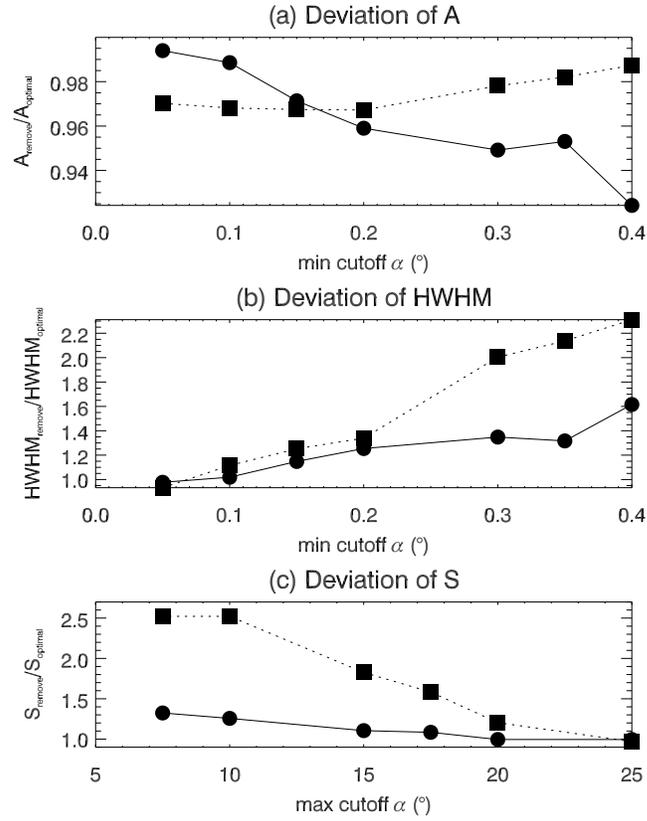}
\caption{\label{fig3} Deviation of the linear-exponential model of \citet{2001JQSRT..70..529K} for the morphological parameters 
A~(a), HWHM~(b) and S~(c) when the phase angle coverage is incomplete. Circles and squares correspond to the cutoff of the 
incomplete phase functions of respectively the B~ring and C~ring typical phase curves (Fig.\,\ref{fig2}).}
\end{center}
\end{figure}

\textbf{[Fig.\,\ref{fig2}, Fig.\,\ref{fig3}]}

In Fig.\,\ref{fig3}, we represent the ratio of the morphological 
parameters of the incomplete phase curve over the morphological parameters of the 
fuller phase curve, called $A_{\textrm{remove}}/A_{\textrm{optimal}}$, 
HWHM$_{\textrm{remove}}$/HWHM$_{\textrm{optimal}}$ and $S_{\textrm{remove}}/S_{\textrm{optimal}}$. 
We observe a slight underestimation of A, a strong overestimation of HWHM 
and a moderate underestimation of S. \newline 
The deviation of the optimal value is quite weak for A ($A_{\textrm{remove}}/A_{\textrm{optimal}}\thicksim$0.96 
at a cutoff of 0.3\textsuperscript{o}) but its variation depends on the morphology of the surge. 
For the typical B ring phase curve (which has a narrower peak), we note a slight decrease 
of the ratio $A_{\textrm{remove}}/A_{\textrm{optimal}}$ and for the C~ring, we note a 
slight increase of the ratio. However the both ratios lead to an underestimation, which 
means that incomplete data fitted by the morphological model will have a smaller 
amplitude that data which cover the full surge. \newline
HWHM shows the strongest deviation of $A_{\textrm{remove}}$. Indeed, HWHM is 
overestimated for the typical B ring phase curve (HWHM$_{\textrm{remove}}$/HWHM$_{\textrm{optimal}}\thicksim$1.4  
at a cutoff of 0.3\textsuperscript{o}) but is strongly overestimated for the typical C~ring 
(HWHM$_{\textrm{remove}}$/HWHM$_{\textrm{optimal}}\thicksim$2.0 at a cutoff of 
0.3\textsuperscript{o}) which has a wider peak. \newline
For the slope, we notice an overestimation in the order of  $S_{\textrm{remove}}/S_{\textrm{optimal}}\thicksim$1.3 
at a cutoff of 7\textsuperscript{o} for the B ring phase curve and of $\thicksim$2.5 
for the C ring at the same cutoff. This means that the slope of the linear part is 
stabilized at roughly 15\textsuperscript{o}. \newline
These comparisons are important when we compare morphological trends found by \citet{2002Icar..158..224P}, in section\,\ref{comparative_approach} and \citet{2007PASP..119..623F}, in section\,\ref{compare_hst_iss}.

\subsubsection{Linking morphological parameters with the physical parameters of the models} \label{phys_morph}
The use of a simple morphological model is generally not adapted to derive the 
physical properties of the medium. However, the theories developed for the 
coherent backscattering and the shadow hiding effects deduce their properties 
by parameterizing the opposition phase curve 
\citep{1992MNRAS.254P..15M,1992Ap&SS.189..151M,1992Ap&SS.194..327M,1999Icar..141..132S,1986Icar...67..264H,2002Icar..157..523H}. 
Recent laboratory experiments raise some doubts on the meaning of the physical parameters of these models 
and their correlation with the real physical properties of the medium \citep{2007JGRE..11203001S,2007JQSRT.106..487S}.
However, because there is no better modelization, we connect the morphological parameters 
A, HWHM and S with the physical characteristics of the medium derived from these models. 
\begin{description}
\item[A] is the amplitude of the opposition peak and describes the behavior of the phase function at the 
smallest phase angles ($\alpha$<2\textsuperscript{o}). What we know about the opposition effect is that it occurs 
at the smallest phase angles and acts on the multiple scattered light in
the regolith on the surface of the particles~: the underlying phenomenon is the coherent
backscattering effect. The coherent-backscattering effect increases in brightness by
almost a factor two, while using grains size smaller 
than the wavelength of the incident light, \citep{1992Ap&SS.189..151M,1999Icar..141..132S}. In contrast, the
second phenomenon of the opposition effect, the shadow hiding effect,
is known to produce a wide peak from 0 to 2 degrees, and to decrease
the brightness up to 20 degrees \citep{1986Icar...67..264H,S1999}. The combination of the two 
effects at very low phase angle is still a matter of debate and today two theories disagree in order to explain
the peak of the opposition. The theory of \citet{1992Ap&SS.194..327M,1992Ap&SS.189..151M} 
assumes that the opposition peak is a pure coherent backscattering effect whereas the theory of \citet{2002Icar..157..523H} shows that the opposition peak results from a coupling of coherent-backscattering and shadow
hiding, even at low phase angles. This coupling should be due to the fact
that the coherent backscattering could act on both multiple and single
scattered light whereas the shadow hiding is a single scattered light
effect \citep{2002Icar..157..523H}. Thus, this theoretical model defines two amplitudes 
as output parameters~: the coherent backscatter amplitude~$B_{C0}$ and the shadow hiding 
amplitude~$B_{S0}$. As a consequence, using this theory, it does not seem to be
possible to ascribe the morphological parameter A solely to the coherent backscattering.  \newline
For most laboratory measurements \citep{1999Icar..141..132S,2000Icar..147..545N}, 
the amplitude of the opposition peak is a function of grain size in such way that 
A decreases with increasing grain size. 
This anti-correlation finds a natural explanation by the fact that macroscopic irregularities 
(>$\lambda$) create less coherent effects than microscopic irregularities ($\leq\lambda$). 
Laboratory measurements of \citet{2003AA...409..765K} also confirmed that the opposition surge 
increases when irregularities are small. \citet{1992Ap&SS.189..151M} and \citet{1992Ap&SS.194..327M} underline the fact 
A is linked to the intensity of the background $I_{b}$ which is a decreasing 
function of increasing absorption \citep{1990AdSpR..10..187L}, thus A must 
increase with increasing absorption or decreasing albedo~$\varpi_{0}$, which 
was also confirmed by laboratory measurements of \citet{2003AA...409..765K} who found that the peak
decreases with increasing sample albedo. 
\item[HWHM], the half-width at half-maximum, is generally associated to the 
coherent backscatter effect. It can be related to the grain size, index of 
refraction and packing density of regolith \citep{1992Ap&SS.194..327M,1992MNRAS.254P..15M,2002Icar..157..523H}. 
The HWHM is maximum for a effective grain size near $\lambda$/2 and increases 
when the regolith grains filling factor~$f$ increases. For high values of $f$, the 
HWHM shifts towards the greater grain size. However, 
as for the amplitude, the model of \citet{2002Icar..157..523H} defines two HWHMs~: 
the coherent backscatter angular width ($h_{c}$), which is defined similarly that in the model of 
\citet{1992Ap&SS.194..327M}, and the shadow hiding angular width ($h_{s}$). This reinforces the idea that the 
observed surge results from a coupling between the coherent backscattering and the shadow hiding.
\item[S], the slope of the linear part, seems to be due only to the
shadow hiding effect: the interferences caused by
coherent backscattering effect seem not to be very significant at
larger phase angles \citep{M2006,2002Icar..157..523H}. This means
the shadow hiding light is not affected by the coherent backscattering at larger phase angles. \newline
Also, according to recent analytical and numerical models, the shadow hiding acts solely on 
the linear part of the phase function \citep{S1999,1999Icar..141..132S}. We define here the absolute slope, whereas
it is naturally a negative parameter, because the phase function always decreases in brightness from 
10\textsuperscript{o} to 40\textsuperscript{o} of phase angles \citep{1974IAUS...65..441K,S1999,2007PASP..119..623F}. \newline
The slope depends on the volume filling factor~D and the optical depth of the slab \citep{1974IAUS...65..441K,S1999}. 
Two models has been developed for different regimes of the particle volume density~D and predict 
opposite behavior of S as the function of D and $\tau$. \newline
The \textit{shadowing} model of \citet{1966JGR....71.2931I} and \citet{1974IAUS...65..441K}, 
also called \textit{inter-particle shadow hiding} by \citet{1978Icar...34..227G},  
consists in shadows of macroscopic particles \citep[$\bar{r}$=15~m, see][]{1974IAUS...65..441K} in a particulate medium, 
such as 8D$\ll$1 \citep{1966JGR....71.2931I}. 
The smaller the volume density~D, the steeper the phase function for increasing phase angle \citep{1974IAUS...65..441K}. 
Other refinements of this model exists \citep{1979Icar...39...69E,1986Icar...67..264H,S1999,2007PASP..119..623F} but lead to the same results~:  
the opposition peak due to the shadowing sharpens and the absolute slope increases with decreasing packing density (or filling factor). \newline
Another model exists, the \textit{intra-particle shadow hiding}, which is valid for higher particle volume density according to  \citet{1978Icar...34..227G,1994IAUS..160..271M}. \citet{1985Icar...64..320B} underlined the fact that the mutual shadowing among regolithic grains could be suited for understanding the textural properties of the regolith. As a consequence, this mechanism operates at the surface of ring material~: e.g. in the regolith layer.  \newline
However, other physical parameters need to be taken into account. In the analytical \textit{inter-particle shadow hiding} model of \citet{2002Icar..157..523H}, 
the slope can be linked to the angular width of the shadow hiding~$h_{sh}$. Thus a normalized and absolute slope would be $S=\frac{1}{2h_{sh}}=\frac{\textrm{D}}{Q_{\textrm{ext}}(\lambda,\bar{r})\bar{r}\ln(1-\textrm{D})}$, 
with D the volume filling factor, $Q_{\textrm{ext}}(\lambda,\bar{r})$ the extinction coefficient and 
$\bar{r}$ the mean radius of particles. Then, the slope should depend on the wavelength and on the particle size. 
Interestingly, laboratory experiments of \citet{2003AA...409..765K} showed that the slope increases when the sample's size increases. However, the microscopic and/or macroscopic roughness of the medium need also to be taken into 
account in the shadow hiding models, as underlined by \citet{1984Icar...59...41H,1986Icar...67..264H,1999Icar..141..132S,2003AA...409..765K}
and \citet{2007JGRE..11203001S}. For example, laboratory experiments of \citet{2003AA...409..765K} showed that 
the slope increases when the sample's roughness increases. \newline
Then laboratory measurements and recent theoretical models could significantly increase the number of physical parameters on which S depends.
\end{description} 

From the above arguments, the HWHM and the amplitude A are 
governed by both coherent backscatter and shadow hiding effects whereas 
the slope S provides information of 
the shadow hiding effect solely.   

\section{Shape of the phase curves at opposition} \label{results}
\subsection{The opposition effect in CLEAR filters}  \label{oe_saturn_rings}
Due to the automation of extraction and fitting procedures (cf
section\,\ref{data_reduction}) and due to the high images resolution, phase functions were extracted 
in as many as 211~different locations, in the D, C, B~rings, Cassini Division and A~ring (in
increasing distance from Saturn, left column of Table~1 of Electronic supplementary material). In this section, we first present
the typical behavior of some selected phase curves in different
regions of the main rings (section\,\ref{phase_curves_oe}), then we
discuss similarities and differences and what are the general trends
from sections\,\ref{comparative_approach} to \ref{compare_parameters_oe}.

\subsubsection{Overview of the CLEAR phase curves}   \label{phase_curves_oe}

\begin{figure}[!ht]
\includegraphics[width=15cm]{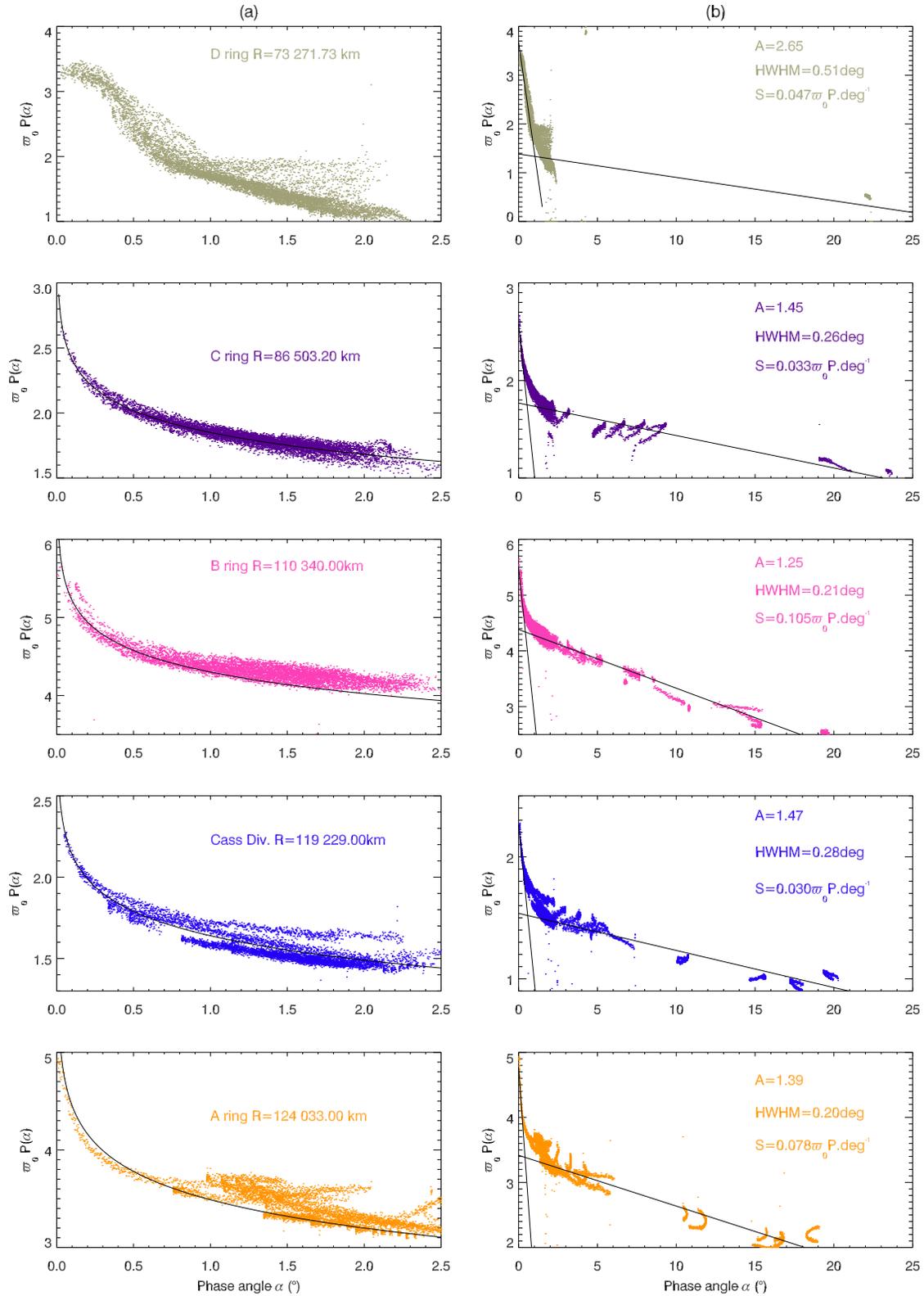}
\caption{\label{represent_oe_allrings} Representative CLEAR phase curves for
the main rings with a zoom on the surge (a) and the full phase curve
(b), fitted respectively with the logarithmic model of \citet{1970sips.conf..376B} (a) and the linear-by-part model of \citet{1976AJ.....81..865L} (b).}
\end{figure}

Examples of phase curves in various ring regions are presented in 
Fig.\,\ref{represent_oe_allrings}. Each pair of graphs show on the
left side a zoom from 0.01 to 2.5 degrees and on the right side, the fuller
phase curve from 0 to 25 degrees. These curves were obtained by
combining several WAC images with a large distribution of viewing
geometries (Table\,\ref{table_oe_info_images}) each curve is built
from the merging of 10 to 70 different images with various values of
emission angle, incidence angle, phase angle etc. The dispersion of points is not due to 
the measurement uncertainty but reflects mainly the limits of the
\citet{1960ratr.book.....C} inversion formula that was used to extract the
phase function from the measured values of I/F. It seems that some
important physics may be missing (like multiple scattering), that could explain the scattering of points.
Curves for the A and D rings (Fig.\,\ref{represent_oe_allrings}) are
incomplete between 3 to 20\textsuperscript{o} which is due to removal
of images because of an artifact (cf section\,\ref{data_set}).  

\textbf{[Fig.\,\ref{represent_oe_allrings}]}

Whereas the general shape is similar from one ring to another (Fig.\,\ref{represent_oe_allrings}), 
some details in the shape may 
vary significantly.
First pair of graphs in Fig.\,\ref{represent_oe_allrings} shows
the phase curve derived in the D~ring from images of
\textit{June~26}. Due to short exposure time (10~ms), the D~ring
ringlets are too faint to be detected. In images of \textit{June~26},
a bright spot is visible from 67~000~km to the inner boundary of the
C~ring~: this corresponds to the expected location of the background
sheet of material constituting part of the D~ring \citep{2007Icar..188...89H}. 
However camera artifact may be visible in such dim regions
of the image but could not be clearly identified here. So the fact
that a strong increase of brightness at the expected location of the
opposition and the coherent variation of the signal with observing
geometry between different images suggest that we indeed see the
opposition effect in the D~ring. However some doubts still
remain. From 0.5 to 2~degrees, the curve is similar to other
rings. Below 0.5 degree an exponential surge and a flattening at zero
degree distinguish this phase function from the other ones. Does it
reflect optical properties of D~ring dust~? Is it an artifact~? This
plateau below 0.5 degree is much too large to be explained by the
finite angular radius of the Sun (0.025 degree). Because of these uncertainties, at this point it is
speculative to interpret this specific behavior as real. 
     
For the C~ring (Fig.\,\ref{represent_oe_allrings}), the shape of the
phase function is well sampled below 2 degrees. It is comparatively
wider than in dense A and B~rings. More precisely, HWHM of the
opposition surge is wider for the C~ring
(HWHM=0.26\textsuperscript{o}) than for the A and B ring
(HWHM$\thicksim$0.20\textsuperscript{o}). This could be also interpreted as
a steeper slope of the linear regime of the phase function for
$\alpha>2$\textsuperscript{o}. Wavy features between 5 to 25 degrees
can be attributed to images artifact. Their amplitude is about 15\% of
the total signal of the C~ring.

The B~ring opposition surge has the
smallest amplitude of all rings (A=1.25 in Fig.\,\ref{represent_oe_allrings}). This was already underlined
by the recent study of \citet{2007PASP..119..623F} whereas previous observations of \citet{1979AJ.....84.1408E} and \citet{2002Icar..158..224P} give the exact opposite trend. 
However, their result could have a bias due to the lack of data below 0.3\textsuperscript{o} 
of phase angle for \citet{2002Icar..158..224P} and 0.1\textsuperscript{o} for \citet{1979AJ.....84.1408E}
whereas \citet{2007PASP..119..623F} had values much smaller \citep[the smallest phase angles at which HST observed the B ring
range from 0.0037 to 0.0132 degrees][]{2007PASP..119..623F}. 
The B~ring has also the steepest slope (S=0.105~$\varpi_{0}P$.deg$^{-1}$) in the linear regime
explaining why the opposition spot is so contrasty in the ISS images.

The Cassini Division has a similar amplitude and width (A$\thicksim$1.47 and HWHM$\thicksim$0.28\textsuperscript{o} in Fig.\,\ref{represent_oe_allrings}) to the C~ring (A$\thicksim$1.45 and HWHM$\thicksim$0.26\textsuperscript{o} in Fig.\,\ref{represent_oe_allrings}). Their slope of the linear regime are also similar (S=0.033~$\varpi_{0}P$.deg$^{-1}$ 
for the Cassini Division and S=0.030~$\varpi_{0}P$.deg$^{-1}$ for the C ring). These similarities were first 
noticed by \citet{2002Icar..158..224P} and are suggestive of a strong dependence
of the opposition effect on the optical depth (we will come back to 
this in section\,\ref{compare_parameters_oe}).

An example of the A ring phase function is given in the last pair of
graphs in Fig.\,\ref{represent_oe_allrings}. At first sight, the
signal appears much more disturbed than in other rings~: specifically,
pieces of phase functions extracted from different images show a wide
dispersion in this graph, whereas signal from an individual image has
a very low dispersion. The origin of this is not clear and may be due
to the artifact reported in section~\ref{observe_et_reduit}. The
dispersion of the data is about 15\% of the signal whereas the camera
artifact should represent at most 5\% of the signal only (estimated on
the background). It may be possible that the dispersion may be also
due to an intrinsic photometric effect which is not corrected by the
\citet{1960ratr.book.....C} single scattering model
(equation(\ref{eq_chandrasekhar})). Indeed, the A~ring has an
intermediate optical depth $\thicksim$0.5 so that it neither appears
as a solid surface (like the B ring) or as a dilute system (like the
C~ring).  Here, we are in an intermediate regime where many collective
effects may influence strongly the apparent phase function (multiple
scattering, gravitational wakes, density waves, etc.). A sophisticated model
is required here to investigate such effect \citep{PW2007}. However the general 
trends are quite clear and the A~ring phase
curve has a larger peak amplitude (A=1.39 in Fig.\,\ref{represent_oe_allrings}) than the B~ring's 
(A=1.25 in Fig.\,\ref{represent_oe_allrings}). We see in
addition that the slope of the linear regime is shallower in the A~ring (S=0.078~$\varpi_{0}P$.deg$^{-1}$ in Fig.\,\ref{represent_oe_allrings}) than in the
B~ring (S=0.105~$\varpi_{0}P$.deg$^{-1}$ in Fig.\,\ref{represent_oe_allrings}) but steeper than in less dense rings (S$\thicksim$0.03~$\varpi_{0}P$.deg$^{-1}$ in Fig.\,\ref{represent_oe_allrings}). Thus the phase curve at
opposition in the A~ring is somewhat intermediate between the B and
C~rings, strengthening the idea of a dependence on the optical depth.

Finally, we conclude this section by remarking that the opposition
effect is very diverse in Saturn's rings, and could be the consequence of 
different properties of the surface ring particles in various
ring regions. Some general trends can be underlined, as we see in the
next section. \newline
In a first step, we check if some correlations exist between the morphological parameters which
depend on the same portion of the curve (e.g. A and HWHM for the surge) and
also if parameters describing different parts of the phase curve can be
correlated (e.g. A, HWHM and S for, respectively, the surge and the linear
part). This is the purpose of the next section.

\subsubsection{Cross comparisons between morphological parameters}
\label{comparative_approach} 
In order to constrain the morphology of the surge, we correlate the amplitude A
with the angular width HWHM for the different main rings. \newline
We fit A and HWHM with a linear function and found correlation coefficients reported 
in Table\,\ref{cross_study_kaasalainen_slope_ahwhm}.
 
\textbf{[Table\,\ref{cross_study_kaasalainen_slope_ahwhm}]}

In Table\,\ref{cross_study_kaasalainen_slope_ahwhm} the Cassini Division, 
the amplitude A has the steepest function of the HWHM. A linear fit
gives a slope of about 1.2 with a correlation coefficient of 47~\%. This mean that
narrow surges have a low amplitude and inversely wide surges have large
amplitude. \newline 
In the C~ring, values of A and HWHM are generally 
greater than in the Cassini Division (but there are some exceptions). For the amplitude, this qualitative 
difference between the Cassini Division and the C~ring is confirmed by 
previous study of \citet{2002Icar..158..224P} and \citet{2007PASP..119..623F}. 
For the angular width, although our trend of HWHM agrees with the results of 
\citet{2007PASP..119..623F}, it contradicts previous work of \citet{2002Icar..158..224P}, 
this is certainly due to the 
overestimation of HWHM when the smallest phase angles are missing 
(section\,\ref{incomplete_data_fits} and Fig.\,\ref{fig3}b). \newline
For the C ring, we find a slope for A=f(HWHM) of about 0.9 with a good
correlation coefficient of 79~\%. \newline 
The A~ring shows a similar
variation of A=f(HWHM) as for the C~ring (1.0) with a correlation
coefficient of 56~\%.  \newline 
Finally, A and HWHM in the B~ring have
values smaller than in the faint rings (C~ring and Cassini
Division). Data points of A and HWHM for the B~ring are concentrated
in a similar range as the A~ring one, however with a much shallower
slope (0.6 with a correlation coefficient weakly reliable of 29~\%,
see Table\,\ref{cross_study_kaasalainen_slope_ahwhm}). This means that
the shape of opposition phase curves in the B~ring may have various
angular width with an almost constant value of the amplitude. \newline
To conclude, the amplitude of the surge seems correlated with the
angular width, at least for the C~ring, A~ring and Cassini Division
whereas the amplitude is independent of HWHM for the B~ring. The slope
of A=f(HWHM) seems to decrease from the Cassini Division to the
B~ring, passing by the C and A~rings, suggesting that the slope is a
decreasing function of the optical depth.  

Whereas the slope~S on the one hand, A and HWHM on the other hand are thought to be related to
different portion of the phase curve (linear part and surge
respectively), it is interesting to note that they are somewhat
correlated. We simply note that S is a decreasing linear function the
angular width and is also correlated with the amplitude. Slopes and correlation coefficients of 
S=f(A) and S=f(HWHM) are reported in Table\,\ref{cross_study_kaasalainen_slope_ahwhm}. This could 
be due to the fact that we derive our slope from 1.5\textsuperscript{o} to 25\textsuperscript{o} 
whereas analytical model of \citet{1999Icar..141..132S}, for example, describes the shadow hiding 
effect as a slope which fits the phase curve from 4.5\textsuperscript{o} to larger phase angles, 
see \citep{2002Icar..158..224P}.

\subsubsection{Regional behavior} \label{subregional_oe} 
\begin{figure}[!ht]
\includegraphics[width=15cm]{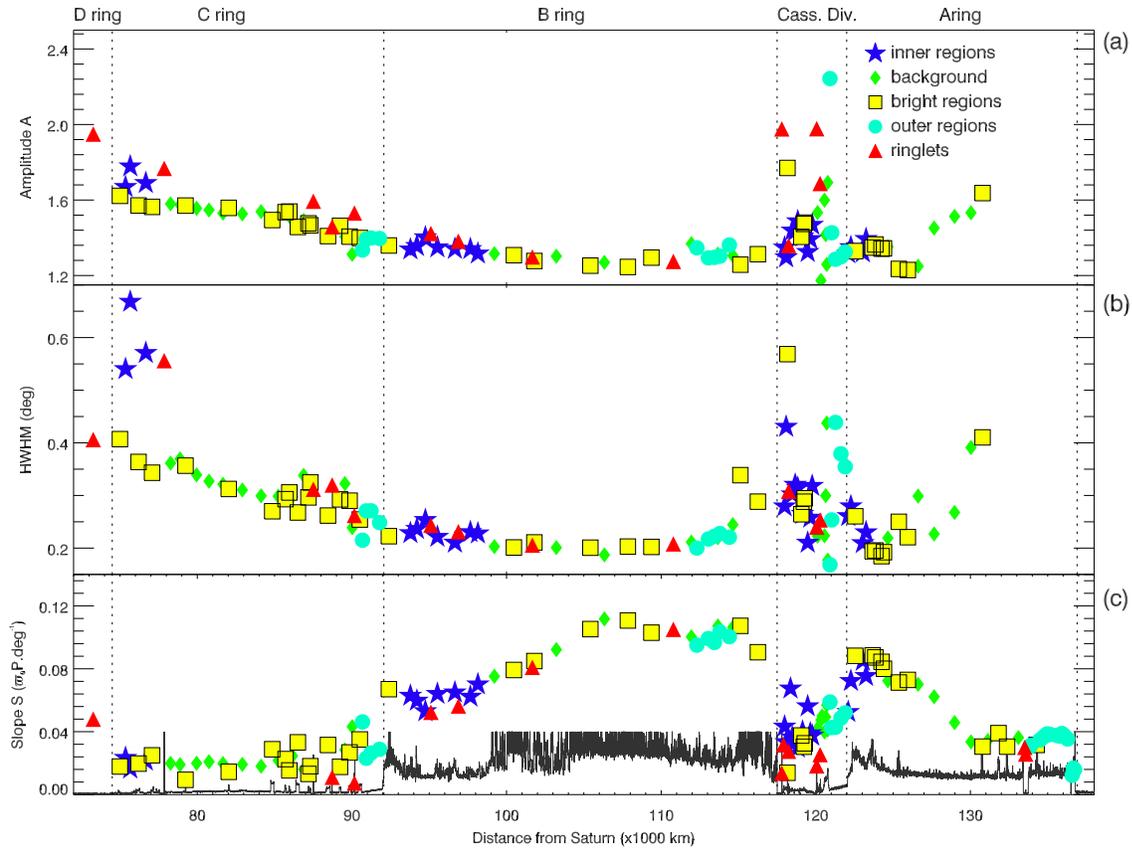}
\caption{\label{ahwhms_ring} Regional behavior of the morphological
parameters~: (a) the amplitude A, (b) the angular width HWHM in degrees~ and (c) the absolute slope S in $\varpi_{0}P$.deg$^{-1}$ from the Linear-by-part model of \citet{1976AJ.....81..865L} for CLEAR phase curves using the ring type classification. 
The vertical dotted lines correspond to ring boundaries. The optical depth of PPS/Voyager is plotted 
in (c) as a radial reference.}
\end{figure}

We now
look for regional behavior inside each ring (Fig.\,\ref{ahwhms_ring}). 
The Fig.\,\ref{ahwhms_ring} displays A, HWHM and
S as a function of the distance from Saturn by introducing a \textit{ring type} nomenclature based on
the regional behavior of the C~ring, well studied by \citet{1991PhDT.........8C} and
decomposed into three \textit{ring types}~: inner ring, background and
\textit{plateaux}. We have modified and extended this nomenclature to
five classes of ring features, then applicable to the entire main
rings system~:
\begin{itemize} 
\item[(1)] inner regions characterized by low optical depth in all the
rings (for example, the dark bands in the Cassini~Division, see \citep{1989Icar...82..180F}),
\item[(2)] background are morphological smooth regions without abrupt
variation of $\tau$,
\item[(3)] bright regions (\textit{plateaux} or \textit{plateaus} in the C~ring \citep{1982Natur.297..115H}, density and bending
waves in the A~ring located by \citet{1983Icar...56..439E}) are
the regions in each ring with the highest optical depth,
\item[(4)] \textit{ringlets} according to \citet{1982Natur.297..115H}
are thinner ring embedded in a less dense region or a gap,
\item[(5)] outer regions (for example the so-called \textit{ramp} for the
C~ring and the Cassini Division \citep{1998Icar..132....1C}) mark the
transition at the boundaries of each ring.
\end{itemize} 

\textbf{[Fig.\,\ref{ahwhms_ring}]}

In Fig.\,\ref{ahwhms_ring}a and b, the amplitude~A and the HWHM vary 
smoothly across the main ring system (C to A rings), with only little scattering around the main 
trend (appart from the Cassini Division), illustrating the stability of the linear-by-part model for
comparing multiple observations in different ring regions. From the C~ring to the middle of the B~ring, 
a smooth decrease is observed. No
sharp transition is observed between the C and B~rings. The outer
regions of the C~ring which are rich in gaps, plateaux and ringlets,
have a somewhat larger value of amplitude. \newline 
From the middle of the B ring to the outer of the A ring, both A and HWHM increase
again. The Cassini Division presents (1) larger values of A and HWHM
as in the C~ring and (2) strongly dispersed values which may be indeed
real because no image artifact is visible in this Division. \newline 
The slope has a significantly different behavior
(Fig.\,\ref{ahwhms_ring}c) because strong jumps are observed at the
boundaries of each ring. This reinforces differences of behavior of
the surge and the linear part of the phase curve with the distance
from Saturn.  

As a result, the behavior of A, HWHM ans S did not show significant variations with ring type. 
\citet{1991PhDT.........8C} noticed single scattering albedo of the C ring was dependent on the ring 
type classification. Maybe that the absence of correlation between morphological 
parameters and the ring type classification implies that the morphological parameters are independent of 
the single scattering albedo. Since the single scattering albedo is correlated with the optical depth, 
\citep{2005EM&P...96..149S}, we try now to correlate the morphological 
parameters with the optical depth in the next section.

\subsubsection{Optical depth variations of the morphological parameters} \label{compare_parameters_oe}
\begin{figure}[!ht]
\begin{center}
\includegraphics[width=6cm]{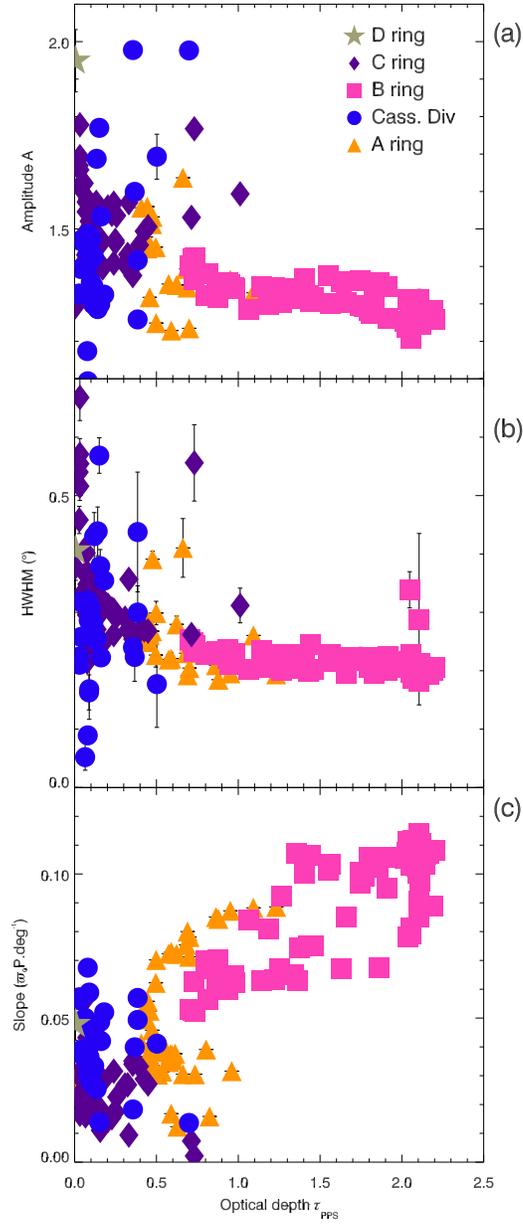}
\caption{\label{results_kaasalainen_tau} Morphological parameters of CLEAR phase curves from the Linear-by-part
model of \citet{1976AJ.....81..865L}~: Amplitude A~(a), Angular width HWHM~(b) and absolute slope
S~(c) in $\varpi_{0}P$.deg$^{-1}$.}
\end{center}
\end{figure}

In order to quantify differences in terms of morphological shape in
our 211~phase curves in CLEAR filters (see
section\,\ref{phase_curves_oe}), we use the three parameters A, HWHM
and S. They are represented in the
Fig.\,\ref{results_kaasalainen_tau} as a function of the normal
optical depth of the rings. 

\textbf{[Fig.\,\ref{results_kaasalainen_tau}]}

Because the overall correlations of A, and S HWHM with $\tau$ are not clear, although they seem to lead to an negative correlation for A and HWHM (Fig.\,\ref{results_kaasalainen_tau}a,b) and a positive correlation for S (Fig.\,\ref{results_kaasalainen_tau}c), we calculated with a linear fit the correlation coefficients for A=f($\tau$), HWHM=f($\tau$) and S=f($\tau$) for each main ring, see Table\,\ref{correlation_tau_ahwhms}.

\textbf{[Table\,\ref{correlation_tau_ahwhms}]}

The amplitude~A of the surge (Fig.\,\ref{results_kaasalainen_tau}a)
is correlated with the optical depth of the rings. The following
trends may be noted~: first, the amplitudes of low optical depth ($\tau$<0.5, typically the C ring and the Cassini Division) are positively correlated with the optical depth (Table\,\ref{correlation_tau_ahwhms}) and second, the amplitude at high optical depth
($\tau$>1, typically the A and B rings) are negatively correlated with the optical depth (Table\,\ref{correlation_tau_ahwhms}). The second trend was quite clear by eye, whereas the first one is more difficult to distinguish in Fig.\,\ref{results_kaasalainen_tau}a. This is due to the fact that the C ring and the Cassini Division have an optical depth restricted in the range 0.01<$\tau$<0.5, which is quite compressed in the scale from 0 to 2.5 of Fig.\,\ref{results_kaasalainen_tau}. The same behavior \citep[increasing of amplitude with increasing albedo, or $\tau$, the two values are correlated][]{1989Icar...80..104D,dones1993} was noticed by \citet{2000Icar..147...94B} for the amplitude of asteroids' phase curves~: for albedo<0.3 an increase of the amplitude with increasing albedo was noticed. \newline
We note also that for intermediate and high optical depth (0.5<$\tau$<2.5, typically the A and B rings) the amplitude 
has a much smaller scattering and finally, the correlation coefficient are larger (-74\% for the B~ring). \newline
For the A~ring, it is interesting to note that regions of lower optical
depth (0.3<$\tau$<0.5) connect well with data-points in the C~ring
both in terms in mean value and scattering (Fig.\,\ref{results_kaasalainen_tau}a). A good continuity with the
B~ring is also observed (0.7<$\tau$<1.1).  

The angular half width of the peak at half maximum (Fig.\,\ref{results_kaasalainen_tau}b)
show similar correlation woth the optical depth than the amplitde of the surge~: increase of HWHM when the optical depth increases for the C~ring and the Cassini Division and decrease of HWHM when the optical depth increases for the A and B~rings. The
scattering of HWHM (HWHM$\thicksim0.3\pm0.2$) for the low optical depth regions behaves similarly as for the
amplitude, and lead to small correlation coefficients (19\%, see Table\,\ref{correlation_tau_ahwhms}). 
As for the amplitude, the scattering of HWHM is narrow for
intermediate optical depth (0.5<$\tau$<1.6). Again, the behavior of
the A~ring is clearly intermediate between the C and B~rings. To
summarize, the behavior of HWHM is a decreasing function of increasing
optical depth, with important scattering at low~$\tau$ that could be understood as an opposite behavior of HWHM (HWHM is an increasing function of increasing optical depth). 

The general trend for the slope of the linear regime
(Fig.\,\ref{results_kaasalainen_tau}c) is a strong increase with
increasing optical depth, with a uniform scattering and with central
value well represented by S$\thicksim0.07\tau^{1/2}$. This is the first time that a 
correlation is established between the slope~S and the optical depth. Previous morphological study on 
asteroids' solar phase curves showed an exponential correlation between the slope of the phase function 
(the so-called phase coefficient~$\beta$ in mag.deg$^{-1}$) and the albedo \citep{2000Icar..147...94B}. 
Our trend for the slope (the increase of S in $\varpi_{0}P$.deg$^{-1}$ with increasing $\tau$) 
is thus consistent with the slope of asteroids' phase functions 
\citep[decrease of $\beta$ in mag.deg$^{-1}$ with increasing albedo, recall that the magnitude~$M$ is inversely proportional to $I/F$, the so-called geometric albedo~: $I/F=10^{-0.4M}$, see][so decrease of $\beta$ in mag.deg$^{-1}$ with increasing albedo leads to increase of $\beta$ in I/F.deg$^{-1}$ with increasing albedo]{1995Icar..115..228D}, since we assume that the optical depth is positively correlated 
with the albedo, as already noticed by \citet{1989Icar...80..104D,dones1993}. However, as for A and HWHM, low optical depth regions (as the C ring and the Cassini Division) have a distinct trend than the trend of the A and B~rings.

In conclusion, whereas the slope S has a strong tendency with the optical depth (the average of the absolute correlation coefficients of S=f($\tau$, see values of Table\,\ref{correlation_tau_ahwhms}) is 53\%), the first two parameters A and HWHM have a soft tendency with the optical
depth (the average of the absolute correlation coefficients for A=f($\tau$) is 44\% and for HWHM=f($\tau$) is 32\%). Consequently, Fig.\,\ref{results_kaasalainen_tau} which presents A, HWHM and S according to the optical depth, yields
the following trends~:
\begin{itemize} 
\item[(1)] The morphological parameters of the surge (A and HWHM) have both similar behavior with $\tau$ in the C~ring and the Cassini Division. Firstly, a positive correlation of A with $\tau$ and of HWHM with $\tau$, with a strong scattering and become almost independent of the optical depth for $\tau$>0.5. Secondly, we note  a negative correlation of S with $\tau$. These trends are reversed (negative correlations of A and HWHM with $\tau$ and a positive correlation of S with $\tau$) for the moderate and high optical depth regions (typically the A and B rings)~; 
\item[(2)] the morphological parameter of the linear regime (S) is strongly positively correlated with the optical depth~: negatively correlated in the C~ring and the Cassini Division (where $\tau$<0.5) and positively correlated in the A and B~rings (where $\tau$>0.5).
\end{itemize} 
We conclude that the trends of all the morphological
parameters are linked to the optical depth. Then the drastic differences between, on the one hand, the amplitude and
the angular width of the surge and, on the other hand, the slope of
the linear part across the main ring system suggests that these characteristics
originate from different physical mechanisms, as predicted by physical models (see section\,\ref{phys_morph}).

\subsection{The opposition effect in COLOR filters}  \label{oe_saturn_rings_color}
\subsubsection{Overview of the COLOR phase curves}   \label{phase_curves_oe_color}

\begin{figure}[!ht]
\includegraphics[width=15cm]{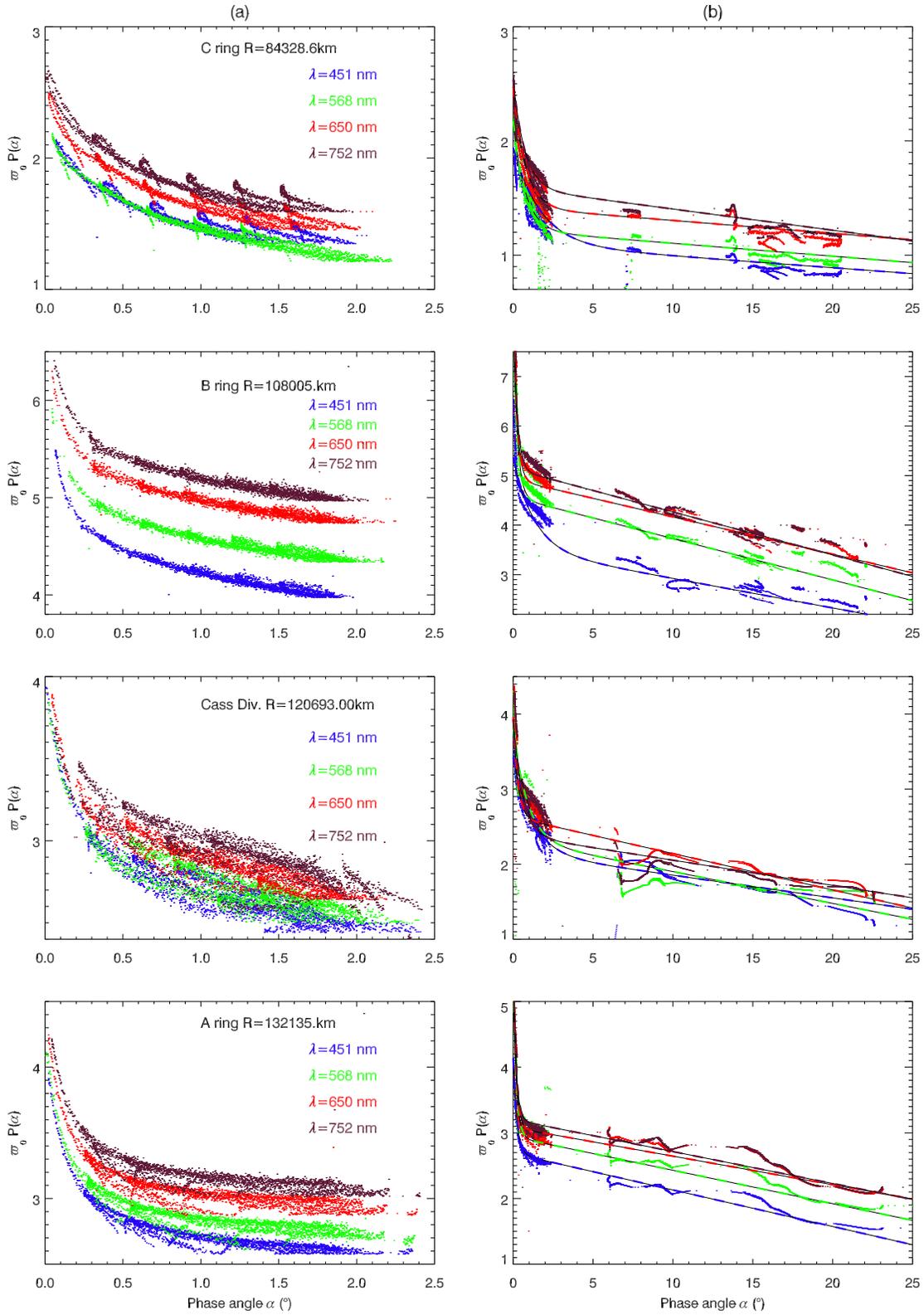}
\caption{\label{represent_oe_allrings_color} Representative COLOR phase curves for
the main rings with a zoom on the surge with WAC images (a) and the full phase curve with NAC and WAC images (b). 
Full phase curves of (b) are fitted with the linear-exponential model of \citet{2001JQSRT..70..529K}.}
\end{figure}

Fig.\,\ref{represent_oe_allrings_color} 
details examples of phase curves obtained for the C, B, A~rings and the Cassini Division. 
A striking result is that the COLOR phase curves' peaks are   
much narrower than the CLEAR phase curves' peaks. Indeed, this is due to the fact that WAC images 
(exclusively used for the CLEAR phase curves) do not have the more peaked part of the phase function 
($\alpha$>0.01\textsuperscript{o}) whereas NAC images have it ($\alpha$>0.001\textsuperscript{o}, 
see Fig.\,\ref{represent_nac_all_rings}). To check this, we plot in the first column (labeled~a) 
of Fig.\,\ref{represent_oe_allrings_color} the phase function obtained with the camera WAC and in the 
second (labeled~b) we plot the fuller phase function obtained with all the NAC and WAC images combined. 
Indeed, Fig.\,\ref{represent_oe_allrings_color} demonstrates that the value of $\varpi_{0}P(\alpha)$ 
when $\alpha $ tend towards 0\textsuperscript{o} is greater in the panel~(b) than in the panel~(a). 

\textbf{[Fig.\,\ref{represent_oe_allrings_color}]}

The fact that WAC images do not include the peakest part of the opposition surge explains also 
the difference between the peak's intensity of curves in Fig.\,\ref{oe_images_color_location} 
in CLEAR filters and COLOR filters because when we processed images in CLEAR filters, there was no NAC images. 
This explains also the shift in I/F at the minimum phase angle on NAC images and WAC images of the same filter  
(Fig.\,\ref{oe_images_color_location}). We observe that the shift in I/F between NAC and WAC image is minimum in 
the blue filter (Fig.\,\ref{oe_images_color_location}), 
which implies that the full surge is contained in WAC images for this wavelength. 

In general, we noticed that the general shape of the curve is similar to that obtained previously~: 
the C~ring still has a fairly broad peak with a large amplitude (first pair 
of graphs at the top of Fig.\,\ref{represent_oe_allrings_color}). 
The B~ring exhibits also narrow peaks as those of CLEAR filters comparatively to the other main rings 
(the second pair of graphs of  Fig.\,\ref{represent_oe_allrings_color}).
The Cassini division (third pair of graphs of Fig.\,\ref{represent_oe_allrings_color}) 
shows again a lot of scattering, 
which could be the consequence of the failure of the \citet{1960ratr.book.....C} inversion 
when the optical depth is not well known.
Finally, the A~ring, yet very dense, shows also scattering for $\varpi_{0}P(\alpha)$, 
which could be due to the gravitational wakes, 
see the last pair of graphs in Fig.\,\ref{represent_oe_allrings_color}. 
Indeed, the \citet{1960ratr.book.....C} inversion is very sensitive to the 
optical depth, and the wakes are known to modify locally the value of $\tau$ 
\citep{2006GeoRL..3307201C,2007AJ....133.2624H}.

\subsubsection{Results in COLOR filters and comparisons with the morphological behaviors of the CLEAR filters}
\label{results_morph_clear_color}

We now consider the variation of morphological parameters A($\lambda$), HWHM($\lambda$) 
and S($\lambda$) in CLEAR and COLOR filters with the distance from Saturn (see Fig.\,\ref{ahwhms_rings_clear_color}).

\textbf{[Fig.\,\ref{ahwhms_rings_clear_color}]}

The amplitude of the opposition in the outer parts of the 
C and A rings has a much broader scattering in COLOR filters that in CLEAR filters 
(Fig.\,\ref{ahwhms_rings_clear_color}a). There is, 
unfortunately, not enough radial coverage to generalize this effect across the C ring. 
The Cassini Division also shows similar behavior for 
the dataset at high and low spectral resolution. 
In the B ring, where there is a good radial coverage in COLOR filters, 
amplitudes of CLEAR filters are much lower than the smallest amplitudes in COLOR 
(typically in the blue at $\lambda$=0.451~$\mu$m), also noticed 
for the amplitudes in CLEAR and COLOR filters in 
the A ring. This could be due to the 
exclusive use of the images of the WAC to resolve the CLEAR opposition surge 
(section\,\ref{phase_curves_oe_color}) \newline
The half-width at half maximum of the peak shows a good agreement 
between the data in CLEAR filters and COLOR filters (Fig.\,\ref{ahwhms_rings_clear_color}b). 
The values are the same order of magnitude in all the rings and the regional effect is the same, 
except for the A ring where HWHM at high spectral resolution (COLOR filters) decreases 
when the distance at Saturn increases while at low spectral resolution (CLEAR filters), HWHM 
seems to start to grow. \newline
Finally, for the slope $S$ of the linear part of the phase function, we 
represent a normalized slope in order to compare data from the CLEAR filters 
with data from the COLOR filters. We follow the method of \citet{2007PASP..119..623F} 
which divided the slope by the intensity of the background, thus the normalized slope 
is $S/B_{1}$ for the linear-by-part model and now has the deg$^{-1}$ unit. 
In Fig.\,\ref{ahwhms_rings_clear_color}c we have for 
$S$ similar behavior as a function from Saturn in the inner A ring and the Cassini Division for data 
in CLEAR filters and in COLOR filters. Finally, the B ring shows regional effects very different 
at high and low spectral resolution, especially in the middle of the B~ring, at 107~000~km 
(the most optically thick region) where the CLEAR filters' slope is significantly 
overestimated compared to the slope in the blue filter ($\lambda$=0.451~$\mu$m). In the same region, we also
observed a strong scattering of HWHM (Fig.\,\ref{ahwhms_rings_clear_color}b).

\subsubsection{Comparisons with multi-wavelength HST observations}   \label{compare_hst_iss}

\begin{figure}[!ht]
\includegraphics[width=15cm]{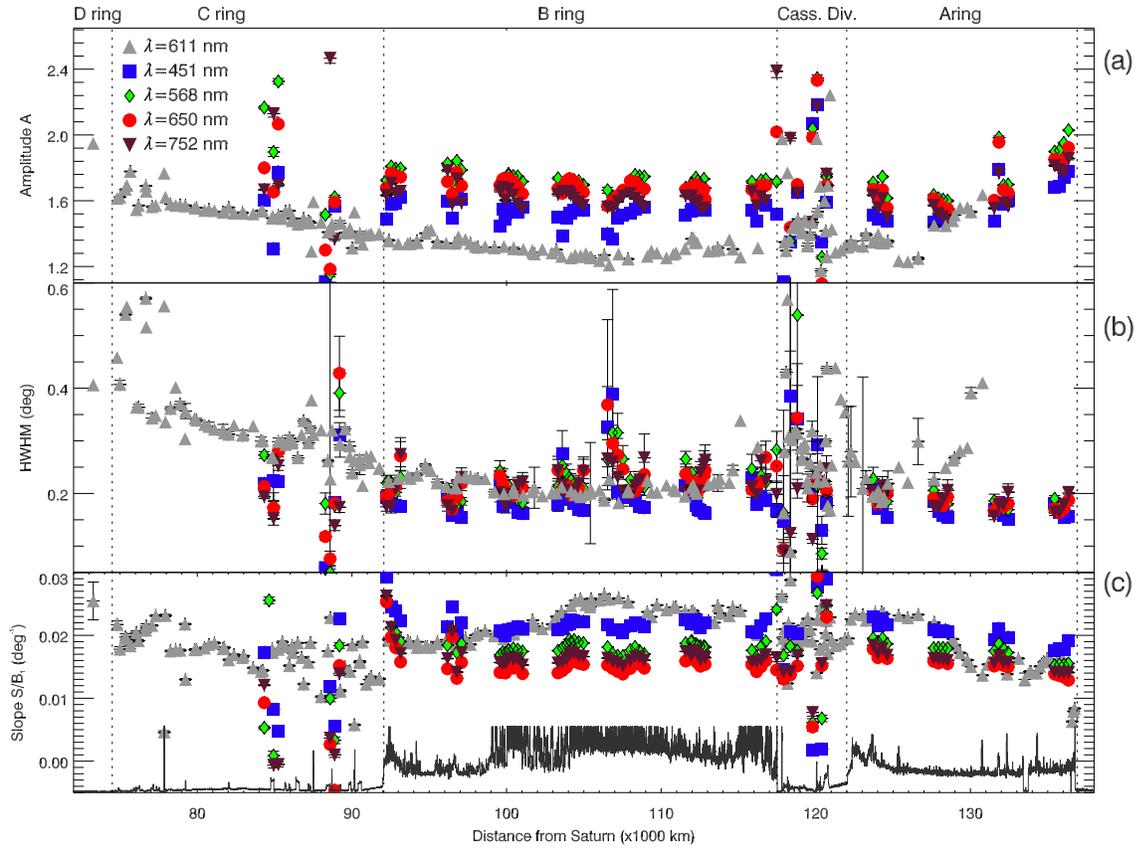}
\caption{\label{ahwhms_rings_clear_color} Regional behavior of morphological
parameters~: (a) the amplitude A, (b) the angular width HWHM and (c) unitless absolute slope $S/B_{1}$ in deg$^{-1}$ from the Linear-by-part model of \citet{1976AJ.....81..865L} 
using CLEAR and COLOR phase curves. The vertical dotted lines correspond to ring boundaries. 
The optical depth of PPS/Voyager is plotted in (c) as a radial reference.}
\end{figure}

We now compare the Cassini data (Fig.\,\ref{ahwhms_rings_clear_color}) 
with Earth-based data. For this purpose, the recent 
study of \citet{2007PASP..119..623F} was chosen due to their small phase angles 
$\alpha$>0.028\textsuperscript{o} and one point below the minimum phase angle 
corresponding to the angular size of the Sun. 
The phase curves of \citet{2007PASP..119..623F} were obtained in I/F for the main rings and adjusted 
with the linear-by-part model of \citet{2001JQSRT..70..529K} which provided morphological 
parameters~A, HWHM and S for different wavelengths ranging from the ultraviolet to infrared (see their figure~7).

We consider first the regional effects of morphological parameters derived by 
\citet{2007PASP..119..623F} with the WFPC2/HST instrument. \newline
The dispersion observed by the HST for the three morphological parameters A, 
HWHM and S in the C ring and the Cassini Division is also very clear with ISS 
(Fig.\,\ref{ahwhms_rings_clear_color}abc). However, 
it should be noted for the C ring scattering observed with our data (especially 
those in CLEAR filters) is highly localized (internal and external parts of the ring) 
and not present in the central regions, corresponding to the background according 
to the ring type classification of \citet{1991PhDT.........8C}. \newline
In the B ring, \citet{2007PASP..119..623F} noticed strong variations of A and S in the inner edge 
which are absent for A and are present for S in the ISS data (Fig.\,\ref{ahwhms_rings_clear_color}a and c). 
For HWHM, HST did not 
have the raise of HWHM in the middle of the ring that we observed (Fig.\,\ref{ahwhms_rings_clear_color}b).
For the slope, we observe in both data set a decrease of S from the inner edge to the 
middle of the ring (Fig.\,\ref{ahwhms_rings_clear_color}c). \newline
In the A ring, the regional trends for the amplitude~A and the angular width HWHM are similar for HST in ISS 
(Fig.\,\ref{ahwhms_rings_clear_color}a), but it is not the case for S. 
\citet{2007PASP..119..623F} obtained an almost constant value of S. 
The absolute value of the slope of \citet{2007PASP..119..623F} in all filters is roughly  
S$\thicksim$0.04~deg$^{-1}$. With ISS/Cassini (Fig.\,\ref{ahwhms_rings_clear_color}c), we observe two 
distincts trends. First, S at high and low spectral resolution have similar values. Second, the regional 
effect observed is the same~: there is a decreasing trend with increasing distance from Saturn. At the inner edge, S$\thicksim$0.025~deg$^{-1}$ 
and at the outer edge S$\thicksim$0.015~deg$^{-1}$. There is, however, 
a different order of magnitude obtained by ISS/Cassini and WFPC2/HST which could be 
due to lack of data at large phase angles for Earth-based observation.
Indeed, the factor 2 between the slope values is 
well explained by the overestimation of the slope when the phase angle coverage stops 
at 6 degrees (section\,\ref{incomplete_data_fits} and Fig.\,\ref{fig3}c).

Now, we turn on the comparison on the variations of the morphological parameters 
with the wavelength, which leads to the following conclusions. \newline
First, \citet{2007PASP..119..623F} found weak variations of A($\lambda$) in the 
B ring. Indeed, figure~3 of \citet{2007PASP..119..623F} shows weak wavelength-variations of the amplitude in the B ring, 
and also in the A ring. Only one notable difference occurs in the ultraviolet. 
Overall, \citet{2007PASP..119..623F} observed a decrease of A($\lambda$) from 
the ultraviolet to the green, and then an increase in the amplitude of the green to the 
infrared. This is not at all what is observed for the amplitudes of ISS where a 
notable increase from blue to green and a decrease in the infrared to the green 
(Fig.\,\ref{ahwhms_rings_clear_color}a). The values of A($\lambda$) for ISS/Cassini and WFPC2/HST are anti-correlated. \newline
For HWHM($\lambda $), \citet{2007PASP..119..623F} remarked a strong decrease of 
the half-width at half maximum when the wavelength increases, and in all the rings. 
No break in decreasing HWHM($\lambda$) is observed by the HST. 
This decrease is very different of the behavior of 
HWHM($\lambda$) with ISS/Cassini (Fig.\,\ref{ahwhms_rings_clear_color}b). Thus the variations of 
HWHM($\lambda$) presented here are also unprecedented. There is firstly not a decrease but an increase of HWHM  
with increasing wavelength. Second, the order of magnitude 
found is not the same. The angular half-width obtained with WFPC2/HST are generally 
between 0.05\textsuperscript{o} and 0.2\textsuperscript{o} while those of ISS 
are between 0.1\textsuperscript{o} and 0.4\textsuperscript{o}. This discrepancy could be the consequence of the 
relative thickness of our phase functions due to the use of several images at similar phase angles 
whereas \citet{2007PASP..119..623F} did not have multiple recovered data because they obtained one point per phase angle. \newline
Finally, for S($\lambda $), the wavelength behavior of \citet{2007PASP..119..623F} are consistent 
with those of ISS/Cassini (S decreases from the blue to the red, and slightly increases in the infrared), 
Fig.\,\ref{ahwhms_rings_clear_color}c. Strangely, this agreement shows a 
consistency of both data sets for variations of the 
slope with a wavelength whereas the regional effects of S observed by the two instruments 
are significantly different.

\subsubsection{Variations of the opposition effect with the wavelength and the optical depth}   

\begin{figure}[!ht]
\begin{center}
\includegraphics[width=6cm]{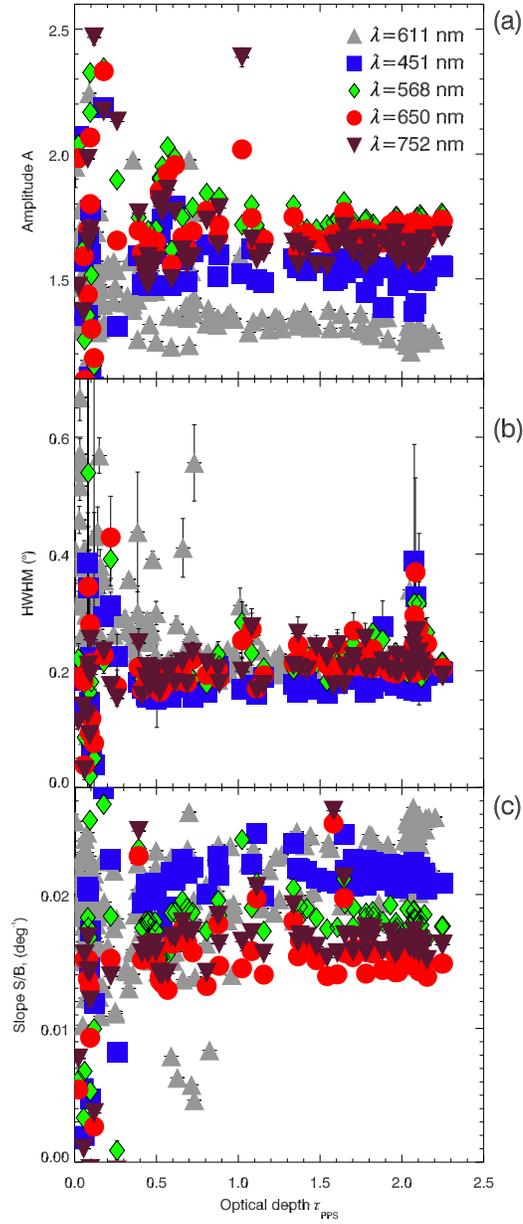}\caption{\label{oe_morph_params_clear_color_tau} Morphological parameters of CLEAR and COLOR  
phase curves from the Linear-by-part
model of \citet{1976AJ.....81..865L}~: (a) the amplitude A, (b) the angular width HWHM and (c) the unitless absolute slope $S/B_{1}$ in deg$^{-1}$.}
\end{center}
\end{figure}

We see that the morphological parameters depend on the optical depth 
(Fig.\,\ref{results_kaasalainen_tau}) and also on the 
wavelength (Fig.\,\ref{oe_morph_params_clear_color_tau}). 

\textbf{[Fig.\,\ref{oe_morph_params_clear_color_tau}]}

\begin{figure}[ht]
\begin{center}
\includegraphics[width=5cm]{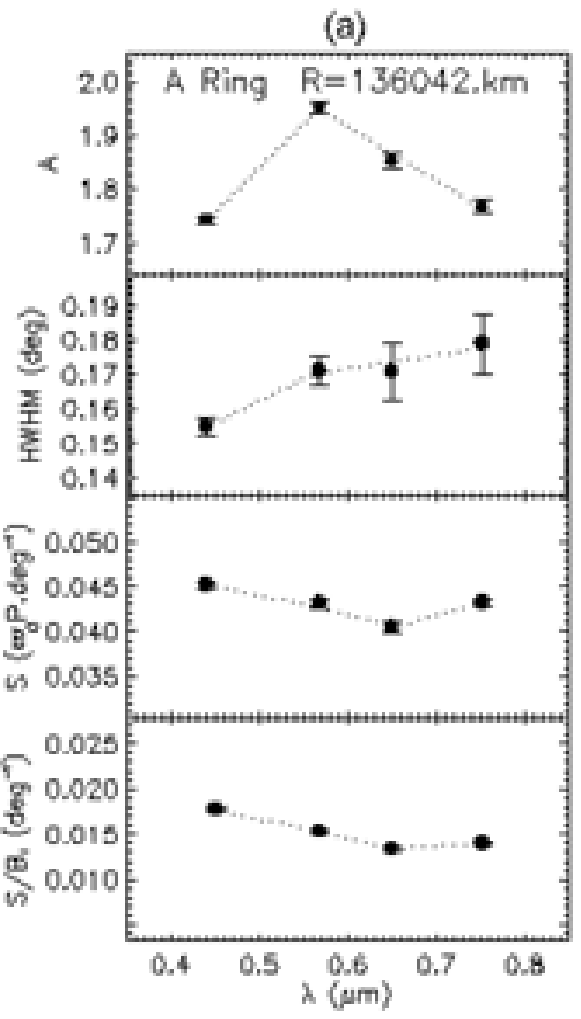}
\includegraphics[width=5cm]{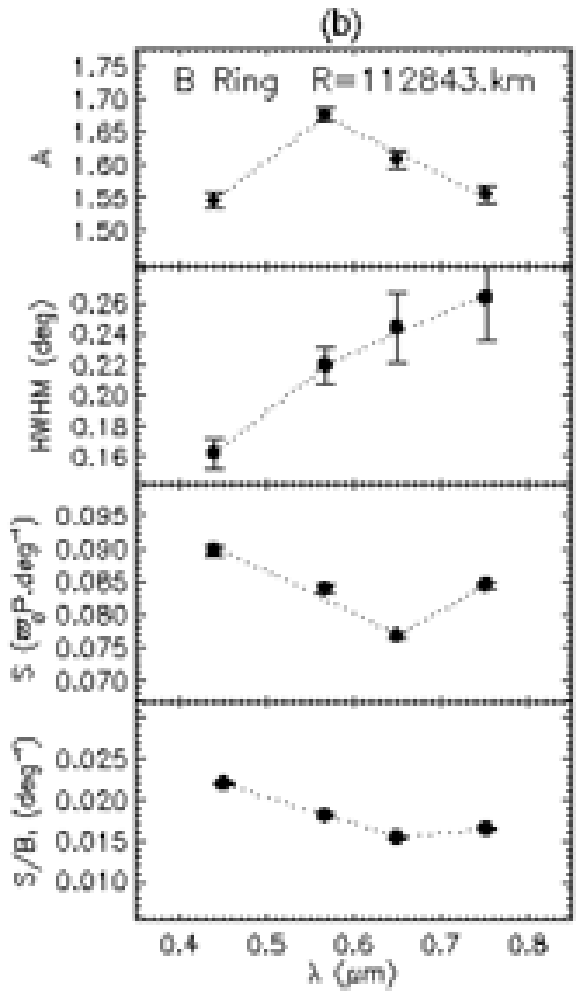}
\end{center}
\caption{\label{linear_by_part_color_rings} \label{lastfig} Variations of the morphological parameters A($\lambda$), HWHM($\lambda$), the absolute slope S($\lambda$) and the unitless absolute slope $S/B_{1}$($\lambda$) for two typical regions of the A ring (a) and in the B ring (b). Dotted lines correspond to linear fits obtained in the spectral ranges. The slopes of these linear functions are called ``steepness'' and are correlated with the rings' optical depth in tables\,\ref{ahwhms_wavelenght_tau} and \ref{s_wavelenght_tau}.}
\end{figure}

It is appropriate now to quantify the variations of A($\lambda$), 
HWHM($\lambda$) and S($\lambda$) by fitting them with a linear model. 
We thus obtain for each COLOR phase curve of the rings two linear functions 
(one from 0.451 to 0.568~$\mu$m and another from 0.568 to 0.752~$\mu$m) 
which fit the behavior of A($\lambda$) and two linear functions HWHM($\lambda$), with the same 
wavelength boundaries. For the 
slope S($\lambda$), two linear functions are obtained (one from 0.451~$\mu$m to 
0.650~$\mu$m and from 0.650~$\mu$m to 0.752~$\mu$m), see Fig.\,\ref{linear_by_part_color_rings} for an example. 

\textbf{[Fig.\,\ref{linear_by_part_color_rings}]}

For this study, is only kept the slope of each linear function (that we called the steepness), that we correlated 
with the optical depth of the rings (the linear functions and their correlation coefficients are given in Tables\,\ref{ahwhms_wavelenght_tau} and \ref{s_wavelenght_tau}). 

\textbf{[Table\,\ref{ahwhms_wavelenght_tau}, Table\,\ref{s_wavelenght_tau}]}

The slopes of A($\lambda$) and HWHM($\lambda$) show both an increase 
with the optical depth from the blue to the green (Table\,\ref{ahwhms_wavelenght_tau}). \newline
From the green to the infrared, we note a decrease of A($\lambda$) with 
$\tau$, but not for HWHM($\lambda$) which can increase or decrease with 
the optical depth.  \newline
However, the strongest trends are found for S($\lambda$) which lead to 
an decrease with $\tau$ of S($\lambda$) from the blue to the red and to 
an increase of S($\lambda$) from the red to the infrared. Values are 
given in Table\,\ref{s_wavelenght_tau}. \newline
It is interesting to note the similar wavelength trends for A and HWHM, 
which are singularly distinct of the wavelength trends of S. This confirms the fact 
that the morphological parameters
originate from different physical mechanisms, as predicted by the physical models (section\,\ref{phys_morph}). 

\subsection{General trends of the opposition phase curves' morphology with the ISS data}
To conclude the section~\ref{results}, the ISS data set
provides several trends of the opposition effect in Saturn's
rings which concern the~:
\begin{itemize} 
\item[(1)] \textbf{regional behavior of A, HWHM and S across the main ring system}~:
indeed, the classification of ring type features defined in
section~\ref{subregional_oe} shows that A and HWHM vary continuously
across the ring system (thus quite independently of the ring type classification), 
whereas S varies abruptly at the boundaries of
each ring (Fig.\,\ref{ahwhms_ring})~;
\item[(2)] \textbf{morphology of the rings' phase curves}~: in
section~\ref{compare_parameters_oe}, strong dependences on the optical depth of the rings of A, HWHM and
S were established~: anti-correlation of A and HWHM with $\tau$ and correlation of S with
$\tau$ (Fig.\,\ref{results_kaasalainen_tau}). A high scattering at low $\tau$ and high $\tau$ 
is noticed for HWHM in CLEAR and COLOR phase curves~;
\item[(3)] \textbf{correlation between the morphological parameters}~: we showed in
the section~\ref{comparative_approach} that the parameters of the
surge A and HWHM are correlated
(Table\,\ref{cross_study_kaasalainen_slope_ahwhm}), whereas there is only a
weak dependence of HWHM and S on the one hand, and A and S on the other hand 
(Table\,\ref{cross_study_kaasalainen_slope_ahwhm})~;
\item[(4)] \textbf{variations of the morphological parameters in CLEAR and COLOR filters}~: 
we remarked in section\,\ref{results_morph_clear_color} that the CLEAR amplitude is smaller than the COLOR ones, because the 
CLEAR data set did not includes NAC images which generally capture the 
highest part of the surge (Fig.\,\ref{ahwhms_rings_clear_color}). 
We showed also that the discrepancy in the slope of \citet{2007PASP..119..623F} is due to 
their phase angle coverages which did not cover the larger phase angles. As a 
consequence, their slopes are overestimated, as predicted by our study on the 
influence of incomplete data (section\,\ref{incomplete_data_fits} and Fig.\,\ref{fig3}c)~;
\item[(5)] \textbf{variations of the morphological parameters with the wavelength and the 
optical depth}~: A($\lambda$) increases from the blue to the green (0.451~$\mu$m-0.568~$\mu$m) and 
decreases from the green to the infrared (0.568~$\mu$m-752~$\mu$m), Fig.\,\ref{oe_morph_params_clear_color_tau}. 
This increase is reinforced when the optical depth increases (Table\,\ref{ahwhms_wavelenght_tau}) 
and is marked by a maximum value of A($\lambda$) in the green regardless 
of the rings. HWHM($\lambda$) also increases from the blue to the green, but no 
general trend is noticed from the green to the infrared (Table\,\ref{ahwhms_wavelenght_tau}). 
S($\lambda$) decreases from the blue to the red (0.451~$\mu$m-650~$\mu$m), then rises from the red to 
the near-infrared (0.650~$\mu$m-752~$\mu$m), Table\,\ref{s_wavelenght_tau}. The wavelength-variation of S is then 
quite distinct to that of the surge parameters (A and HWHM), then suggesting that they may originate from different 
physical mechanisms. 
\end{itemize} 


\section{Discussion}  \label{discussion} 

\subsection{A : a combination of the coherent backscattering with the shadow hiding~?}
We showed in section\,\ref{compare_parameters_oe} that the amplitude of the surge 
had a specific behavior according to the optical depth. With the model of \citet{1999Icar..141..132S}, 
an estimation of the amplitude is possible but refers only to the coherent backscattering
enhancement. \citet{2002Icar..158..224P} derived the following expression
for the amplitude of the surge~: $A\thicksim 1+e^{-d/L}/2$ where $d$
is the effective radius of grains and $L$ is the free mean path of
photons in the regolithic medium. It turns out that the amplitude
could not be greater than 1.5, which contradicts our morphological
results (Fig.\,\ref{ahwhms_rings_clear_color}a). 
This variation might be due to the shadow hiding effect which
is not taken into account in this computation of the amplitude. 

\subsection{Variations of S with the optical depth}
In shadow hiding numerical simulations of \citet{S1999} that take into account the optical depth and the filling factor, the absolute slope of the linear part has roughly the same value when the optical depth is greater 
than 1, whatever the filling factor value. This theoretical prediction is now confirmed 
by our results : we observed a saturation of S at $\tau$>1 (Fig.\,\ref{results_kaasalainen_tau}c). \newline
Also, we observe with the ISS data (Fig.\,\ref{ahwhms_rings_clear_color}c) a wavelength-dependence of 
the slope S. This behavior was predicted by the model of \citet{2002Icar..157..523H} because S depends on 
the extinction coefficient $Q_{\textrm{ext}}(\lambda,\bar{r})$ where $\bar{r}$ is the mean radius of particles
(see section\,\ref{incomplete_data_fits}).


\subsection{Linking photometric behaviors with dynamical concepts}
The present study, thanks to the quality of its data (radial and
angular resolution), allows us to highlight several observational facts
never reported in the history of the observation of the opposition
effect in the Saturn's system. The slope, the angular width and the amplitude of the rings' opposition 
phase curves are clearly correlated with the optical 
depth of the rings (Fig.\,\ref{results_kaasalainen_tau}). 
Whereas a physical description of such a dependence would need a full new physical model, we 
provide here some arguments explaining how the optical depth may indeed 
influence these three parameters. 

The optical depth is both a measure of the local volume filling factor of material, and
the local collisional activity. Indeed, basic analytical computation shows 
that the filling factor of ring particle is proportional to $\tau/H$ (with 
$\tau$ standing for the optical depth and $H$ standing for the vertical 
height of the rings). So one may expect regions of higher optical depth to 
have a much higher filling factor of 
particles. This is also in agreement with local simulation of ring dynamics 
\citep{1988AJ.....95..925W}, that shows that the  vertical width of material 
increases with decrease optical depth, because of the lower efficiency of 
collisional damping. So one may expect the filling factor of particles to 
be an increasing function of the optical depth. 

However, in the case the slope of 
the phase function for phase angles >1\textsuperscript{o}, analytical and numerical models 
\citep{1966JGR....71.2931I,1974IAUS...65..441K,S1999} predict that
S~should depend on the particle filling factor~D and the vertical extension of the layer of particles in such way that the steeper is the slope, the lower is the filling factor. 
Our morphological trends (Fig.\,\ref{results_kaasalainen_tau}c) 
seem unambiguously to contradict the theoretical trends of these models because 
our highest slopes are found in the high optical depth regions. Because high optical depth regions are known to have 
the highest filling factor \citep{2003Icar..164..428S}, this implies that assumptions of diluted layers \citep[D$\lesssim$0.3 and 8D$\ll$1 respectively for][]{S1999,1974IAUS...65..441K} are not suited to the Saturn's rings. \newline
As a consequence, the shadow hiding observed in the Saturn's rings may result of \textit{intra-particle shadow hiding}, as stated by \citet{1985Icar...64..320B}. Another possibility is that the shadow hiding that operates in the rings are maintained in a ``regime of transparency'' of particles \citep{1987BAAS...19..850L,1988BAAS...20..853I} that dominates the shadowing effect due to packing density. 

We now turn to the case of the HWHM and the amplitude. Whereas 
there is still a debate on what determine their value 
\citep{1992Ap&SS.194..327M,1992MNRAS.254P..15M,1992Ap&SS.189..151M,1999Icar..141..132S,2002Icar..157..523H}, 
authors agree to link them the coherent-backscattering process, which may be controlled by the 
regolith at the surface of ring particles \citep{2002Icar..158..224P}. Like for planetary surfaces, the regolith is 
expected to be the result of space fractionation and erosional processes at the surface of the ring-particles, due (in 
particular) to the on-going collisional activity inside rings. Numerical studies of the dynamics of the ring-particles 
have shown that the optical depth is a key parameter controlling the collisional activity of rings.
On the one hand, the number of collisions per orbit per particle is proportional to $\tau$ \citep[in the regime of 
low optical depth , see][]{1988AJ.....95..925W}, on the other hand, the random velocity in a ring of thickness H is 
about $H\times\Omega$ (with $\Omega$ standing for the local keplerian frequency). 
Since $H$ is a decreasing function of $\tau$, thus 
impact velocities are lower in regions of high optical depth.
In short, particles in low-optical depth regions may suffer rare but violent collisions, conversely, in high-optical 
depth regions particles suffer frequent but gentle collisions. \newline
This may explain qualitatively why the HWHM and 
amplitude have different behavior in the data (Fig.\,\ref{ahwhms_ring} and \ref{results_kaasalainen_tau}). 
However, impact velocities have a lower bound 
$\thicksim 2r\times\Omega$ (with $r$ standing for the particle's radius) 
due to the keplerian shear across the diameter of a particle. 
This ``shear dominated limit" is reached when the optical depth is high, typically for $\tau$>1. In such high filling factor 
regime the dynamics of collisions is entirely controlled by the keplerian shear rather than by the random impact 
velocities. This may qualitatively explain why values of HWHM and Amplitude seem constant for $\tau$>1 : in this 
regime, the collisional activity being about independent of optical depth, the physical properties of the regolith may 
be about constant which is indeed observed. 

In conclusion, in the absence of self-consistent physical model of the opposition effect, these qualitative arguments 
show that there are good reasons to believe that the optical depth is a key factor determining the opposition 
effect in the ring through two different mechanisms~:  
\begin{itemize} 
\item[(1)] the optical depth may influence the absolute slope~S, assuming that shadow hiding is the preponderant mechanism at phase angles $\alpha$>1\textsuperscript{o}
\item[(2)] the optical depth may controls the HWHM and the amplitude (at phase angles $\alpha$<1\textsuperscript{o}) 
if the structure of the regolith is influenced by the collisional activity.
\end{itemize} 

\subsection{Comparison of the opposition effect in optical light and infrared light}
A comparison of the solar phase curves of ISS/Cassini and the thermal phase curves of 
CIRS/Cassini yields to the following trends~:
\begin{itemize} 
\item[(1)] \citet{2007Icar..191..691A} found a prominent opposition surge for the thermal phase curves of the 
plateaux, well fitted by a logarithmic model. This is also the case of the solar phase curves of the plateaux 
observed by ISS (Fig.\,\ref{ahwhms_ring}a)~;
\item[(2)] \citet{2007Icar..191..691A} found that the thermal phase curves of the \textit{background} (regions in 
the close environment of the plateaux) do not have opposition surge, whereas background has an opposition 
surge in the solar phase curves of ISS (Fig.\,\ref{ahwhms_ring}a)~;
\end{itemize} 

A priori, the emitted phase curves may not reflect the coherent backscatter 
effect \citep{2007Icar..191..691A} because interferences of photons did 
not act on heat, and thus on infrared light. However, a pure shadow hiding 
model such as \citep{1981AJ.....86.1694L} fails to reproduce the CIRS opposition 
surge of the plateaux \citep[in general the shadow hiding models did not produce high surges,][]{S1999}.
This could be the proof that the shadow hiding cannot produce solely the opposition surge in emitted light 
and that the coherent backscatter could act on the 
shadow hiding mechanism, by multiplying the single scattered light component at small 
phase angles, as underlined by \citet{2002Icar..157..523H}. \newline
Interestingly, no similar surge were observed in the background \citep{2007Icar..191..691A}, 
whereas both plateaus and background have 
an opposition surge in the solar phase curves of ISS. Because the 
background regions are more dim and reflective than the plateaus \citep[background have 
smaller optical depth and higher albedo than plateaus which means they reflect more than they 
absorb,][]{1991PhDT.........8C}, this could 
explain why these regions did not have an opposition peak in emitted light.


\section{Conclusion} \label{conclusion} 

We report here the main conclusions of this morphological study on the 
Saturn's rings opposition effect seen by ISS/Cassini~:
\begin{itemize} 
\item[(1)] The amplitude~A and the half angular width~HWHM of the opposition surge decrease with 
increasing optical depth~$\tau$. A and HWHM may reflect both coherent backscattering and 
shadow hiding, because according to \citet{2007PASP..119..623F} the morphological parameters of the surge 
are greater than their coherent backscatter counterparts~;
\item[(2)] All the morphological parameters are linked together. We find correlations between A and HWHM, between HWHM and S, and
also between A and S, which imply that S could be more or less affected by the coherent backscattering. 
This could be due to the fact that we derive our slope from 1.5\textsuperscript{o} to 25\textsuperscript{o} whereas analytical 
model of \citet{1999Icar..141..132S} describes the shadow hiding effect as a slope which fits the phase curve 
from 4.5\textsuperscript{o} to larger phase angles, see \citep{2002Icar..158..224P}. 
\item[(3)] The absolute slope~S of the linear part of the phase function increases with increasing 
optical depth~$\tau$ (for optically thick rings) and shows distinct trends to the morphological surge parameters, 
which implies that this parameter is not totally affected by the coherent backscattering. 
As \citep{1987BAAS...19..850L,1988BAAS...20..853I}, we think that the Saturn's rings could be in a regime of transparency of particles because the effect of packing density (decrease of S with increasing packing density) is not that expected for the Saturn's rings (increase of S with optical depth, and if we assume that the optical depth and the packing density are correlated, increase of S with increasing packing density)~;
\item[(4)] $\tau$-dependence with the morphological parameters strengthen our assumptions 
saying that environmental effects are the key element determining the 
opposition effect because the optical depth is a direct measure of the collisional 
and dynamical activity in the surrounding of particles and is highly correlated with A, HWHM and S~;
\item[(5)] Comparisons of ISS/Cassini solar phase curves and CIRS/Cassini thermal phase curves in the C ring 
show that the C ring's plateaus can have a strong opposition surge both in solar and thermal phase curves whereas 
the C ring's background has a strong opposition surge in the solar phase curve and no opposition surge in 
the thermal phase curve.
\item[(6)] Wavelength variations of the amplitude~A of the surge show a maximum in all 
the rings at $\lambda$=0.568~$\mu$m. The increase of A from 0.451 to 0.568~$\mu$m 
and the decrease of A from 0.568 to 0.752~$\mu$m are reinforced with increasing~$\tau$~; 
\item[(7)] Wavelength variations of HWHM of the surge show also a maximum at 
$\lambda$=0.568~$\mu$m but it is not systematic, HWHM can also increase 
from 0.568~$\mu$m to 0.752~$\mu$m. Moreover, there is no specific wavelength variations 
of HWHM with the optical depth~;
\item[(8)] Wavelength variations of the slope~S of the linear part imply that the 
shadow hiding depend on the wavelength, may be via the particle's scattering 
cross-section of recent model \citep{2002Icar..157..523H}. The decrease of S from 0.451 to 0.650~$\mu$m and 
the increase of S from 0.650 to 0.752~$\mu$m are reinforced with increasing~$\tau$.
\end{itemize} 
The goal of this first paper was not to derive and quantify directly the
physical properties obtained from the models. First, because there is
a large set of models and it seemed more convenient to separate the
morphological models to the more physical and sophisticated
ones. Second, because recent physical models did not implement only the opposition 
effect but the main photometric effects which occur in the full phase function, 
from 0 to 180~degrees. Consequently, more investigations will be
provided for this purpose by using full phase curves and photometric 
analytical models in the second paper. In the future, we hope to have the linear degree of polarization at the opposition to obtain more constrains on the ring particle' textures.

\end{linenumbers}
\label{lastpage}

\newpage
\ack
The authors would like to thank F. Poulet, C. Ferrari, S. Rodriguez and the two referees for useful comments 
that improved the quality of the paper. This work was supported by the French \textit{Conseil R/'egional de la Martinique}, the French \textit{Centre National de Recherche Scientifique} (CNRS) and the Cassini Project.

\newpage
\bibliographystyle{elsart-harv}
\bibliography{OE1_corrected}


\newpage

\begin{table}[!ht]
\caption{\label{table_oe_info_images} Main observational parameters of each sequence of images for each 
geometry of observation ($i$=arccos$(\mu)$ and $\epsilon$=arccos$(\mu_{0})$). CLEAR (WAC) filters designate the broadband filters in the optical domain (the central wavelength is $\lambda_{\textrm{cl}}^{\textrm{\textsc{wac}}}=0.611\mu$m). COLOR (NAC) filters designate  
blue, green, red and near infrared filters (at central wavelengths of $\lambda_{\textrm{bl}}^{\textrm{\textsc{nac}}}=0.440\mu$m and $\lambda_{\textrm{bl}}^{\textrm{\textsc{nac}}}=0.451\mu$m;  $\lambda_{\textrm{grn}}^{\textrm{\textsc{nac}}}=0.568\mu$m; $\lambda_{\textrm{red}}^{\textrm{\textsc{nac}}}=0.650\mu$m;  $\lambda_{\textrm{ir}}^{\textrm{\textsc{nac}}}=0.752\mu$m)
and COLOR (WAC) filters designate blue, green, red and near infrared filters (at central wavelengths of $\lambda_{\textrm{bl}}^{\textrm{\textsc{wac}}}=0.460\mu$m; $\lambda_{\textrm{grn}}^{\textrm{\textsc{wac}}}=0.567\mu$m; $\lambda_{\textrm{red}}^{\textrm{\textsc{wac}}}=0.649\mu$m; $\lambda_{\textrm{ir}}^{\textrm{\textsc{wac}}}=0.742\mu$m).}
\begin{center}
{\renewcommand{\arraystretch}{1}
\begin{tabular}{|r|l|r|r|r|r|r|}
\hline
date & Nb im &  $i$ & $\epsilon$ & Radial res. & Azim. res. &  Filters (Camera) \\
      &  &  (\textsuperscript{o})    & (\textsuperscript{o})    & \tiny{(km.pix$^{-1}$)} & \tiny{(km.pix$^{-1}$)}   &  \\
\hline
 \textit{June ~7} 2005 & 12 & 111.5 & 111.9 & 44.0 & 115.1 & CLEAR (WAC) \\ 
 \textit{June 26} 2005 & 66 & 111.3 & 111.3 & 30.1 & 70.0  & CLEAR (WAC) \\ 
 \textit{July 23} 2006 & 48 &  73.1 &  73.4 & 13.4 & 40.7  & CLEAR (WAC) \\ 
\hline
 \textit{May 20} 2005  & 57 & 111.6 & 111.6 &  4.6 &  11.5 & COLOR (NAC) \\ 
 \textit{May 20} 2005  & 59 & 111.5 & 111.9 & 44.0 & 115.1 & COLOR (WAC) \\ 
 \textit{Dec 31} 2006  & 12 & 104.4 & 108.7 & 38.4 & 104.8 & COLOR (WAC) \\ 
 \textit{Feb 20} 2007  & 20 & 103.7 & 122.7 & 66.3 &  72.5 & COLOR (WAC) \\ 
 \textit{Apr 25} 2007  & 16 & 102.7 & 110.7 & 44.4 &  92.6 & COLOR (WAC) \\ 
\hline
\end{tabular}}
\end{center}
\end{table}


\newpage
\begin{table}[!ht]
\begin{center}
\caption{\label{cross_study_kaasalainen_slope_ahwhm}  Results of linear
fits (function and correlation coefficient) obtained for A=f(HWHM), S=f(A) and S=f(HWHM) for each Saturn's main ring and using CLEAR phase curves.}
{\renewcommand{\arraystretch}{1}
\begin{tabular}{|l|r|c|r|c|r|c|}
\hline
 & \multicolumn{2}{c|}{A=f(HWHM)} &  \multicolumn{2}{c|}{S=f(A)} & \multicolumn{2}{c|}{S=f(HWHM)} \\
\cline{2-7}
 & function & correl. & function & correl. & function & correl.  \\
\hline
Cass. Div.     &  0.91 + 1.72$\tau$  &  59~\% & 0.04 -0.03$\tau$ & -68~\% & 0.049 -0.05$\tau$ & -35 \%  \\
C ring         &  1.20 + 0.91$\tau$  &  79~\% & 0.04 -0.06$\tau$ & -46~\% & 0.025 -0.04$\tau$ & -47 \%  \\
A ring         &  1.12 + 1.08$\tau$  &  56~\% & 0.10 -0.14$\tau$ & -84~\% & 0.079 -0.24$\tau$ & -81 \%  \\
B ring         &  1.18 + 0.58$\tau$  &  31~\% & 0.15 -0.33$\tau$ & -79~\% & 0.141 -0.81$\tau$ & -65 \%  \\
\hline
\end{tabular}}
\end{center}
\end{table}


\newpage
\begin{table}[!ht]
\begin{center}
\caption{\label{correlation_tau_ahwhms}  Results of linear fits obtained (function and correlation coefficient) for A=f($\tau$), HWHM=f($\tau$) and S=f($\tau$) for each Saturn's main ring and using CLEAR phase curves.}
{\renewcommand{\arraystretch}{1}
\begin{tabular}{|l|c|r|c|r|c|r|}
\hline
 & \multicolumn{2}{c|}{A=f($\tau$)} &  \multicolumn{2}{c|}{HWHM=f($\tau$)} & \multicolumn{2}{c|}{S=f($\tau$)} \\
\cline{2-7}
 & function & correl. & function & correl. & function & correl.  \\
\hline
Cass. Div.     & 1.22 + 0.937   &  41~\% & 0.20 + 0.175$\tau$ &  19~\% & 0.040 - 0.030$\tau$ & -42~\%  \\
C ring         & 1.50 + 0.024   &  25~\% & 0.33 + 0.077$\tau$ &  19~\% & 0.024 - 0.026$\tau$ & -68~\%  \\
A ring         & 1.50 - 0.175   & -36~\% & 0.31 - 0.106$\tau$ & -49~\% & 0.024 + 0.025$\tau$ &  19~\%  \\
B ring         & 1.42 - 0.068   & -74~\% & 0.24 - 0.016$\tau$ & -39~\% & 0.035 + 0.031$\tau$ &  82~\%  \\
\hline
\end{tabular}}
\end{center}
\end{table}

\newpage
\begin{table}[!ht]
\begin{center}
\caption{\label{ahwhms_wavelenght_tau} Results of linear
fits (function and correlation coefficient) obtained for the steepness of A($\lambda$)=f($\tau$) and for the steepness of HWHM($\lambda$)=f($\tau$) using COLOR phase curves of the A and B rings ($\tau$>0.5).}
{\renewcommand{\arraystretch}{1}
\begin{tabular}{|l|c|r|c|r|}
\hline
 & \multicolumn{2}{c|}{Steepness of A($\lambda$)=f($\tau$)} & \multicolumn{2}{c|}{Steepness of HWHM($\lambda$)=f($\tau$)}  \\
\cline{2-5}
 & function & correl. & function & correl.  \\
\hline
0.451~$\mu$m <$\lambda$< 0.568~$\mu$m  & 1.210 + 0.3$\tau$ &  55~\% & 0.14 + 0.011$\tau$ & 38~\% \\
0.568~$\mu$m <$\lambda$< 0.752~$\mu$m  & 0.001 - 0.5$\tau$ & -51~\% & 0.04 - 0.037$\tau$ & -46~\% \\
\hline
\end{tabular}}
\end{center}
\end{table}

\newpage
\begin{table}[!ht]
\begin{center}
\caption{\label{s_wavelenght_tau} \label{lasttable} Results of linear
fits (function and correlation coefficient) obtained for the steepness of S($\lambda$)=f($\tau$) and the steepness of S/B$_{1}$($\lambda$)=f($\tau$) using COLOR phase curves of the A and B rings ($\tau$>0.5).}
{\renewcommand{\arraystretch}{1}
\begin{tabular}{|l|r|r|r|r|}
\hline
 & \multicolumn{2}{c|}{Steepness of S($\lambda$)=f($\tau$)} & \multicolumn{2}{c|}{Steepness of S/B$_{1}$($\lambda$)=f($\tau$)} \\
 & \multicolumn{2}{c|}{(in $\varpi_{0}P$.deg$^{-1}$.$\mu$m$^{-1}$)} & \multicolumn{2}{c|}{(in deg$^{-1}$.$\mu$m$^{-1}$)}  \\
\cline{2-5}
 & function & correl. & function & correl. \\
\hline
0.451~$\mu$m <$\lambda$< 0.650~$\mu$m  & -0.21 - 0.043$\tau$ & -73~\%  & -0.04 - 0.039$\tau$ & -49~\% \\
0.650~$\mu$m <$\lambda$< 0.752~$\mu$m  &  0.04 + 0.018$\tau$ & 67~\%   &  0.02 + 0.054$\tau$ & 44~\%\\
\hline
\end{tabular}}
\end{center}
\end{table}

\newpage

\newpage

\newpage

\newpage

\newpage

\newpage

\newpage

\newpage

\newpage

\newpage

\newpage

\newpage

\newpage



\newpage
\textbf{Electronic supplementary material} 
\newline

%
%

Table~1: Outputs of the logarithmic model of \citet{1970sips.conf..376B} for the CLEAR phase curves 
representing each ring types (\textit{inner} corresponds to inner regions characterized by low optical depth, \textit{background} are morphological smooth regions without abrupt variation of optical depth, \textit{bright} corresponds to bright regions that have the highest optical depth in each ring, \textit{ringlet} corresponds to a thinner ring embedded in a less dense region or a gap, and \textit{outer} corresponds to outer regions that mark the transition at each ring boundary) of each main ring. Horizontal lines correspond to the ring boundaries. We give for each ring type, the normal optical depth of the Voyager/PPS instrument and the corresponding radius (the distance from Saturn's center). f$_{\varpi_{0}P\rightarrow\textrm{I/F}}$ is a conversion factor which corresponds to the mean level of I/F curves over the mean level of $\varpi_{0}P$ curves.
\setlongtables
{\renewcommand{\arraystretch}{1}
\begin{longtable}[c]{|l|l|l|r|r|c|}
\hline
\hline
Rad. (km) & $\tau_{\textrm{\textsc{pps}}}$& Ring type &~~$a_{0}~~$ &~~$a_{1}$~~& ~~f$_{\varpi_{0}P\rightarrow\textrm{I/F}}$~~   \\
\hline
      73271.7 & 0.005 & ringlet & 2.023 & -0.6457 & 0.00001   \\
\hline      
\endfirsthead
\hline
Rad. (km) & $\tau_{\textrm{\textsc{pps}}}$& Ring type &~~$a_{0}~~$ &~~$a_{1}$~~& ~~f$_{\varpi_{0}P\rightarrow\textrm{I/F}}$~~   \\
\hline \endhead
\hline
\multicolumn{6}{|r|}{table continues on next page...}  \\
\hline \endfoot
\hline
\hline \endlastfoot
      74996.2 & 0.058 & bright & 1.105 & -0.2097 & 0.0223   \\
      75356.9 & 0.027 & inner & 1.331 & -0.2857 & 0.1255   \\
      75665.0 & 0.032 & inner & 1.072 & -0.2535 & 0.1232   \\
      76199.8 & 0.132 & bright & 1.128 & -0.1899 & 0.1255   \\
      76671.2 & 0.031 & inner & 1.188 & -0.2561 & 0.1253   \\
      77075.4 & 0.119 & bright & 1.263 & -0.2106 & 0.0643   \\
      77862.3 & 0.731 & ringlet & 0.568 & -0.1074 & 0.0488   \\
      78273.0 & 0.079 & background & 1.242 & -0.2130 & 0.0501   \\
      78889.1 & 0.074 & background & 1.182 & -0.2030 & 0.0545   \\
      79238.2 & 0.331 & bright & 0.807 & -0.1237 & 0.0310   \\
      79956.9 & 0.096 & background & 1.217 & -0.1985 & 0.1189   \\
      80757.8 & 0.109 & background & 1.213 & -0.1962 & 0.1259   \\
      81661.4 & 0.117 & background & 1.186 & -0.1845 & 0.0466   \\
      82031.0 & 0.202 & bright & 0.969 & -0.1540 & 0.0539   \\
      82914.0 & 0.133 & background & 1.242 & -0.1907 & 0.0621   \\
      84125.6 & 0.102 & background & 1.352 & -0.2052 & 0.0557   \\
      84844.4 & 0.425 & bright & 1.707 & -0.2325 & 0.0418   \\
      85234.6 & 0.099 & background & 1.290 & -0.1953 & 0.0432   \\
      85706.9 & 0.256 & bright & 1.335 & -0.1978 & 0.1254   \\
      85953.3 & 0.227 & bright & 1.098 & -0.1625 & 0.1252   \\
      86158.7 & 0.075 & background & 1.506 & -0.2179 & 0.1249   \\
      86503.2 & 0.396 & bright & 1.853 & -0.2430 & 0.1250   \\
      86877.4 & 0.066 & background & 1.404 & -0.1984 & 0.0522   \\
      87189.3 & 0.153 & bright & 0.892 & -0.1213 & 0.0430   \\
      87312.8 & 0.163 & bright & 1.019 & -0.1440 & 0.1140   \\
      87506.6 & 1.011 & ringlet & 0.645 & -0.0566 & 0.1163   \\
      88451.6 & 0.239 & bright & 1.869 & -0.2257 & 0.1241   \\
      88725.6 & 0.156 & ringlet & 1.087 & -0.1390 & 0.1223   \\
      89233.7 & 0.248 & bright & 1.695 & -0.2038 & 0.1221   \\
      89547.1 & 0.045 & background & 1.995 & -0.2505 & 0.1232   \\
      89851.1 & 0.307 & bright & 1.970 & -0.2265 & 0.1233   \\
      90019.4 & 0.073 & background & 2.361 & -0.2384 & 0.0687   \\
      90163.1 & 0.713 & ringlet & 1.373 & -0.1719 & 0.0633   \\
      90509.7 & 0.355 & bright & 2.310 & -0.2514 & 0.0895   \\
      90685.9 & 0.076 & outer & 2.454 & -0.2532 & 0.1144   \\
      90929.2 & 0.099 & outer & 1.981 & -0.2161 & 0.0985   \\
      91237.3 & 0.137 & outer & 2.060 & -0.2253 & 0.1128   \\
      91788.6 & 0.170 & outer & 2.276 & -0.2447 & 0.0857   \\
      92390.9 & 1.629 & bright & 3.548 & -0.3668 & 0.0952   \\
\hline
      93781.3 & 1.359 & inner & 3.398 & -0.3418 & 0.1254   \\
      94201.6 & 0.935 & inner & 3.377 & -0.3479 & 0.1256   \\
      94751.3 & 0.706 & inner & 3.010 & -0.3528 & 0.1253   \\
      95107.0 & 0.732 & ringlet & 2.984 & -0.3582 & 0.1249   \\
      95527.3 & 0.936 & inner & 3.421 & -0.3460 & 0.1252   \\
      96659.0 & 1.340 & inner & 3.637 & -0.3472 & 0.0542   \\
      96885.4 & 0.817 & ringlet & 3.213 & -0.3492 & 0.0608   \\
      97661.4 & 0.976 & inner & 3.413 & -0.3394 & 0.1265   \\
      98146.4 & 0.872 & inner & 3.497 & -0.3387 & 0.1264   \\
      99213.5 & 1.458 & background & 3.721 & -0.3418 & 0.1259   \\
      100475. & 2.057 & bright & 3.882 & -0.3486 & 0.1281   \\
      101671. & 1.188 & ringlet & 3.698 & -0.3279 & 0.0057   \\
      101800. & 2.104 & bright & 3.939 & -0.3401 & 0.1259   \\
      103223. & 1.266 & background & 3.975 & -0.3634 & 0.1236   \\
      105422. & 2.070 & bright & 4.149 & -0.3446 & 0.0818   \\
      106327. & 2.069 & background & 4.139 & -0.3577 & 0.0477   \\
      107847. & 2.116 & bright & 4.302 & -0.3634 & 0.0679   \\
      109367. & 2.053 & bright & 4.428 & -0.4181 & 0.1249   \\
      110789. & 1.815 & ringlet & 4.250 & -0.4146 & 0.1251   \\
      111953. & 1.407 & background & 4.292 & -0.4617 & 0.0766   \\
      112309. & 1.910 & outer & 4.412 & -0.4370 & 0.1249   \\
      113053. & 1.768 & outer & 4.441 & -0.4350 & 0.1253   \\
      113441. & 1.744 & outer & 4.393 & -0.4334 & 0.1263   \\
      113667. & 1.358 & background & 4.306 & -0.4675 & 0.1269   \\
      113796. & 1.563 & outer & 4.394 & -0.4632 & 0.1261   \\
      114411. & 1.746 & outer & 4.344 & -0.4653 & 0.1262   \\
      114637. & 1.440 & background & 4.342 & -0.4818 & 0.1266   \\
      115122. & 2.047 & bright & 4.506 & -0.5140 & 0.1268   \\
      116254. & 2.105 & bright & 4.547 & -0.4917 & 0.1249   \\
\hline
      117803. & 0.698 & ringlet & 0.722 & -0.2112 & 0.1265   \\
      117910. & 0.087 & ringlet & 1.506 & -0.0942 & 0.1266   \\
      117983. & 0.113 & inner & 2.119 & -0.2572 & 0.1272   \\
      118084. & 0.119 & inner & 1.883 & -0.2250 & 0.1269   \\
      118168. & 0.150 & bright & 1.329 & -0.2249 & 0.1288   \\
      118241. & 0.099 & ringlet & 1.324 & -0.1754 & 0.0204   \\
      118365. & 0.080 & inner & 2.172 & -0.1818 & 0.0222   \\
      118482. & 0.091 & inner & 1.960 & -0.2528 & 0.0464   \\
      118668. & 0.079 & inner & 1.897 & -0.2426 & 0.0225   \\
      118836. & 0.083 & inner & 1.718 & -0.2306 & 0.0475   \\
      119061. & 0.082 & bright & 1.878 & -0.2338 & 0.0435   \\
      119145. & 0.089 & inner & 1.958 & -0.2494 & 0.1003   \\
      119229. & 0.104 & bright & 1.640 & -0.2175 & 0.0508   \\
      119285. & 0.097 & bright & 1.824 & -0.2418 & 0.0561   \\
      119476. & 0.029 & inner & 2.604 & -0.3045 & 0.0626   \\
      119644. & 0.038 & inner & 1.986 & -0.2493 & 0.0527   \\
      119768. & 0.032 & inner & 2.186 & -0.2975 & 0.0639   \\
      120060. & 0.356 & ringlet & 1.509 & -0.2411 & 0.0518   \\
      120116. & 0.161 & background & 1.939 & -0.2629 & 0.0560   \\
      120279. & 0.134 & ringlet & 1.600 & -0.2347 & 0.1158   \\
      120335. & 0.077 & background & 1.980 & -0.2603 & 0.0448   \\
      120408. & 0.064 & background & 2.168 & -0.2863 & 0.0423   \\
      120565. & 0.367 & background & 2.328 & -0.3048 & 0.0370   \\
      120638. & 0.385 & background & 2.574 & -0.3377 & 0.0051   \\
      120711. & 0.384 & background & 2.625 & -0.3319 & 0.0326   \\
      120773. & 0.502 & background & 2.621 & -0.3319 & 0.0465   \\
      120918. & 0.089 & outer & 2.637 & -0.3581 & 0.0952   \\
      121031. & 0.116 & outer & 2.318 & -0.2747 & 0.1058   \\
      121272. & 0.140 & outer & 2.427 & -0.2736 & 0.1093   \\
      121626. & 0.156 & outer & 2.685 & -0.2951 & 0.0563   \\
      121901. & 0.180 & outer & 2.786 & -0.3296 & 0.0790   \\
\hline
      122097. & 0.458 & inner & 2.887 & -0.3562 & 0.1254   \\
      122269. & 0.623 & inner & 3.366 & -0.4469 & 0.1253   \\
      122553. & 1.092 & bright & 3.862 & -0.4958 & 0.1247   \\
      123040. & 0.864 & inner & 3.696 & -0.4431 & 0.1254   \\
      123249. & 0.681 & inner & 3.478 & -0.4386 & 0.1235   \\
      123676. & 1.231 & bright & 3.911 & -0.4554 & 0.1254   \\
      123848. & 0.951 & bright & 3.731 & -0.4466 & 0.1251   \\
      124252. & 0.878 & bright & 3.576 & -0.4149 & 0.1238   \\
      124409. & 0.690 & bright & 3.451 & -0.4045 & 0.1246   \\
      124659. & 0.574 & background & 3.305 & -0.3920 & 0.1245   \\
      125367. & 0.699 & bright & 3.075 & -0.3468 & 0.1256   \\
      125951. & 0.590 & bright & 3.212 & -0.3655 & 0.1256   \\
      126619. & 0.498 & background & 3.074 & -0.3872 & 0.1256   \\
      127655. & 0.495 & background & 2.929 & -0.3742 & 0.1257   \\
      128977. & 0.473 & background & 2.694 & -0.3465 & 0.1256   \\
      130037. & 0.476 & background & 2.568 & -0.3344 & 0.1254   \\
      130786. & 0.661 & bright & 2.637 & -0.3575 & 0.1254   \\
      131122. & 0.465 & background & 2.556 & -0.3406 & 0.1254   \\
      131818. & 0.803 & bright & 2.764 & -0.3932 & 0.1253   \\
      132372. & 0.735 & bright & 2.478 & -0.3015 & 0.1253   \\
      133105. & 0.463 & background & 2.438 & -0.3082 & 0.1253   \\
      133513. & 0.012 & ringlet & 3.354 & -1.0378 & 0.1253   \\
      133574. & 0.027 & ringlet & 1.028 & -0.3205 & 0.1251   \\
      133820. & 0.509 & outer & 2.196 & -0.1977 & 0.1250   \\
      133917. & 0.534 & outer & 2.375 & -0.2581 & 0.1250   \\
      134280. & 0.959 & bright & 2.390 & -0.2465 & 0.1249   \\
      134493. & 0.575 & outer & 2.435 & -0.2647 & 0.1249   \\
      134737. & 0.585 & outer & 2.440 & -0.2657 & 0.1247   \\
      134965. & 0.558 & outer & 2.429 & -0.2550 & 0.1246   \\
      135166. & 0.548 & outer & 2.340 & -0.2079 & 0.1245   \\
      135595. & 0.614 & outer & 2.282 & -0.1600 & 0.1247   \\
      135808. & 0.596 & outer & 2.232 & -0.1423 & 0.1246   \\
      135913. & 0.523 & outer & 2.263 & -0.1494 & 0.0234   \\
      136070. & 0.586 & outer & 2.194 & -0.1160 & 0.0526   \\
      136288. & 0.602 & outer & 2.154 & -0.0992 & 0.1254   \\
      136579. & 0.627 & outer & 1.795 & -0.0080 & 0.1216   \\
      136665. & 0.590 & outer & 1.963 & -0.0159 & 0.1206   \\
      136736. & 0.824 & outer & 1.714 & -0.0040 & 0.1198   \\
\hline      
      140338. & 0.065 & ringlet & 0.240 & -0.0090 & 0.1214   \\
\end{longtable}}
\label{res_kaasalainen_typical_rings}

\end{document}